\documentclass[prc,aps,fleqn,twocolumn,twoside]{revtex4} 
\usepackage[dvipdfm]{graphicx}
\usepackage{dcolumn}
\usepackage{bm}
\newcommand{\be}{\begin{equation}}
\newcommand{\ee}{\end{equation}}
\newcommand{\bea}{\begin{eqnarray}}
\newcommand{\eea}{\end{eqnarray}}

\newcommand{\tg}{\tilde{\gamma}}
\newcommand{\tvx}{\tilde{v}_x}
\newcommand{\tvy}{\tilde{v}_y}
\newcommand{\tvh}{\tilde{v}_{\eta}}
\newcommand{\iw}{\frac{1}{\omega}}
\newcommand{\dET}{\frac{d\epsilon}{dT}}
\newcommand{\dEm}{\frac{d\epsilon}{d\mu}}
\newcommand{\dPT}{\frac{dp}{dT}}
\newcommand{\dPm}{\frac{dp}{d\mu}}
\newcommand{\dnT}{\frac{dn_B}{dT}}
\newcommand{\dnm}{\frac{dn_B}{d\mu}}
\newcommand{\vba}{\mbox {\boldmath$b_A$}}
\newcommand{\vbb}{\mbox {\boldmath$b_B$}}
\newcommand{\vs}{\mbox {\boldmath$s$}}
\newcommand{\vecr}{\mbox {\boldmath$r$}}
\newcommand{\vj}{\mbox {\boldmath$j$}}
\newcommand{\vn}{\mbox {\boldmath$n$}}
\newcommand{\vsigma}{\mbox {\boldmath$\sigma$}}
\newcommand{\vPT}{\mbox {\boldmath$P_T$}}

\begin{document}
\bibliographystyle{h-physrev}

\author{Chiho Nonaka}
\affiliation{School of Physics and Astronomy, University of Minnesota, Minneapolis, MN 55455, USA, \\ Department of Physics, Nagoya University, Nagoya 464-8602, Japan}
\author{Steffen A.~Bass}
\affiliation{Department of Physics, Duke University, Durham, NC 27708, USA}

\title{Space-time evolution of bulk QCD matter}

\begin{abstract}
We introduce a combined fully three-dimensional 
macroscopic/microscopic transport approach 
employing relativistic 3D-hydrodynamics for the early, dense, deconfined stage 
of the reaction and a microscopic non-equilibrium model for the later hadronic
stage where the equilibrium assumptions are not valid anymore.
Within this approach we study the dynamics of hot, bulk QCD matter, which
is being created in ultra-relativistic  heavy ion collisions
at RHIC.
Our approach is capable of self-consistently calculating the
freezeout of the hadronic system, while
accounting for the collective flow on the
hadronization hypersurface generated by the QGP expansion.
 In particular,
we perform a detailed analysis of the reaction dynamics,
hadronic freezeout, and transverse flow.
\end{abstract}

\maketitle

\pagebreak

\section{Introduction}

A major goal of colliding heavy-ions at relativistic energies is to heat up a
small region of space-time to temperatures as high as are thought
to have occurred during the
early evolution of the universe, a few microseconds after the big
bang~\cite{KolbTurner}. In ultra-relativistic heavy-ion collisions, such
as are currently being explored at the Relativistic Heavy-Ion Collider (RHIC),
the four-volume of hot and dense matter,
with temperatures  above $\sim150$~MeV, is
on the order of $\sim (10$~fm$)^4$. The state of strongly
interacting matter at such high temperatures (or density of
quanta) is usually called quark-gluon plasma (QGP)~\cite{QGP}.

The first five years of RHIC operations
at  $\sqrt{s_{NN}}=130$~GeV and $\sqrt{s_{NN}}=200$~GeV
have yielded a vast amount of interesting and
sometimes surprising results \cite{rhic_data1,rhic_flow,rhic_hbt}, many of
which have not yet been fully evaluated or understood by theory.
There exists mounting evidence that RHIC has created
a hot and dense state of deconfined QCD matter with properties similar to
that of an ideal fluid \cite{Ludlam:2005gx,Gyulassy:2004zy} -- this state of matter
has been termed the {\em strongly interacting Quark-Gluon-Plasma} (sQGP).

Heavy-Ion collisions at RHIC involve several distinct reaction stages,
starting from the two initial ground states of the colliding nuclei,
followed by the high density phase in which a sQGP is formed
up to the final freeze-out of hadrons.

The central problem in the study of the sQGP is that
the deconfined quanta of a sQGP are not directly
observable due to the fundamental confining property of the
physical QCD vacuum. If we could see free quarks and gluons
(as in ordinary plasmas) it would be trivial to verify the QCD
prediction of the QGP state. However, nature chooses to hide those
constituents within  the confines of color neutral composite many
body systems -- hadrons. One of the main tasks in relativistic
heavy-ion research is to find clear and unambiguous connections
between the transient (partonic) plasma state and the observable hadronic final
state (for a review on QGP signatures, please see \cite{qgprev}).

One particular approach to this problem is the application of 
transport theory. Transport theory ultimately aims at casting the entire
time evolution of the heavy-ion reaction
-- from its initial state to freeze-out --
into one consistent framework. By tuning the physical parameters of
the transport calculation to data one can then infer from these the properties
of the hot and dense QCD matter of the sQGP and compare these
to the predictions made by Lattice Gauge Theory (LGT).

\section{Specific Model for High-Energy Heavy-Ion Collisions}

Relativistic Fluid Dynamics (RFD, see e.g.
\cite{Bjorken:1982qr,Clare:1986qj,Dumitru:1998es}) is ideally suited
for the {\em QGP and hydrodynamic expansion} reaction phase, but
breaks down in the later, dilute, stages of the reaction when the
mean free paths of the hadrons become large and flavor degrees of
freedom are important. The most important advantage  of  RFD is that it
directly incorporates an equation of state as input and thus is so
far the only dynamical model in which a phase transition can
explicitly be incorporated. In the ideal fluid approximation (i.e.
neglecting off-equilibrium effects) -- and once an initial
condition has been specified -- the EoS is the {\em only} input
to the equations of motion and relates directly to properties of the
matter under consideration. The hydrodynamic
description has been very successful
\cite{Kolb:2003dz,Huovinen:2003fa,Hirano:2002ds}
 in describing the collective behavior of soft particle production at 
RHIC.

Conventional RFD calculations need to assume a
{\em freezeout} temperature at which the hydrodynamic evolution is terminated
and a transition from the zero mean-free-path approximation of a
hydrodynamic approach to the infinite mean-free-path of free streaming
particles takes place. The freezeout temperature usually is a free
parameter which (within reasonable constraints) can be fitted to measured
hadron spectra.

The reach of RFD can be extended and the problem of having to terminate
the calculation at a fixed freezeout temperature can be overcome
by combining the RFD calculation
with a microscopic hadronic cascade model -- this kind of hybrid approach
(dubbed {\em hydro plus micro}) was pioneered in \cite{BaDu00}
and has been now also taken up by other groups \cite{TeLaSh01,Hirano:2005xf}.
Its key advantages are that the freezeout now occurs naturally as a
result of the microscopic evolution and that flavor degrees of freedom are
treated explicitly through the hadronic cross sections of the microscopic
transport.
Due to the Boltzmann equation being
the basis of the microscopic calculation in the hadronic phase,
viscous corrections for the hadronic phase are by default
included in the approach.

Here, we combine the hydrodynamic approach with the microscopic Ultra-relativistic
Quantum-Molecular-Dynamics (UrQMD) model \cite{uqmdref1}, 
in order to provide an improved description of 
the later, purely hadronic stages of the reaction.
Such hybrid macro/micro transport calculations  are to date the
most successful approaches
for describing the soft physics at RHIC. The biggest advantage of the RFD
part of the calculation is
that it directly incorporates an equation of state as input - one of its
largest limitations is that it requires thermalized initial conditions
and one is not able to do an ab-initio calculation.

\subsection{Hydrodynamics}

In the present paper we shall use a fully three-dimensional 
hydrodynamic model \cite{nonaka_refs} for the description of RHIC 
physics, especially focusing on Au + Au collisions at RHIC energies
($\sqrt{s_{NN}}=200$ GeV per nucleon-nucleon pair). 
Our original code for solving the hydrodynamic equations,
which  based on Cartesian coordinates \cite{nonaka_refs}, 
has been modified to the description on the coordinate by 
longitudinal proper time $\tau=\sqrt{t^2-z^2}$  and  
$\eta=\frac{1}{2}\ln [(t+z)/(t-z)]$,   
in order to optimize the 
hydrodynamic expressions for ultra-relativistic  heavy-ion collisions.

In hydrodynamic models, the starting point is the  
relativistic hydrodynamic equation
\begin{equation}
\partial_\mu T^{\mu \nu} = 0, 
\label{Eq-rhydro}
\end{equation}
where $T^{\mu \nu}$ is the energy momentum tensor which is given by   
\begin{equation}
T^{\mu \nu}=(\epsilon + p) U^{\mu} U^{\nu} - p g^{\mu \nu}. 
\end{equation} 
Here $\epsilon$, $p$, $U$ and $g^{\mu \nu}$ are energy density, 
pressure, four velocity and metric tensor, respectively.      
We solve the relativistic hydrodynamic equation Eq.\ (\ref{Eq-rhydro})   
numerically with baryon number $n_B$ conservation    
\begin{equation}
\partial_\mu (n_B (T,\mu) U^\mu)=0. 
\end{equation}
In order to rewrite the relativistic hydrodynamic equation Eq.\ (\ref{Eq-rhydro})  
in the coordinate $(\tau, x, y, \eta)$, we introduce the 
following variables 
\cite{Hi01},   
\begin{equation}
\begin{array}{ccl}
\tvx & = & v_x \cosh Y_L/\cosh(Y_L-\eta), \\
\tvy & = & v_y \cosh Y_L/\cosh(Y_L-\eta), \\
\tvh & = & \tanh(Y_L-\eta), 
\end{array}
\end{equation}
where $Y_L=\frac{1}{2} \ln ((1+v_z)/(1-v_z))$, $\eta=1/2 \ln ((t+z)/(t-z))$. 
Equation (\ref{Eq-rhydro}) in the explicit way, is rewritten in the Appendix.  

In order to solve the relativistic hydrodynamic equations, we
adopt Lagrangian hydrodynamics.
In Lagrangian hydrodynamics, the coordinates of the individual cells do
not remain fixed, but
move along the flux of the fluid.
In the absence of turbulence during the expansion, 
Lagrangian hydrodynamics has several advantages over the conventional Eulerian 
approach:
\begin{itemize}
\item computational expediency: a fixed number of cells can be utilized
through the entire calculation.  
A Lagrangian hydrodynamic code can thus easily be employed even at
ultra-high energy collisions such as at the Large Hadron Collider (LHC)  
where a large difference of scale exists between the initial state 
and the final state due to the large gamma factor and rapid expansion 
of the QCD matter.   
\item analysis efficiency: the adiabatic path of each 
volume element of fluid can be traced in the phase diagram, making it 
possible to directly discuss the effects of the phase transition on physical 
observables \cite{nonaka_refs}. 
\end{itemize}

Our algorithm for solving the relativistic hydrodynamic  
equation in 3D is based on the conservation laws for entropy 
and baryon number. 
Further details concerning the numerical method can
be found in ref.\ \cite{nonaka_refs}.  

\subsection{Equation of State}
To solve the relativistic hydrodynamic equation, an equation of 
state (EoS) needs to be specified.  
The inclusion of an equation of state as input is one of
the biggest advantages  of  RFD, which 
is so far the only dynamical model in which a phase
transition can
explicitly be incorporated. In the ideal fluid approximation
(i.e. neglecting off-equilibrium effects),
the EoS is the {\em only}
input to the equations of motion and relates directly to
properties
of the matter under consideration.
Once the EoS has been fixed (e.g. through a lattice-QCD calculation)
a comparison to data can be used to extract information
on the initial conditions of the hydrodynamic calculation \cite{Hu05}.

Lattice-QCD (lQCD) offers the only rigorous approach for
determining the EoS of QCD matter.
Calculations at vanishing baryon chemical potential
 suggest that for physical values of the quark masses (two light ($u,d$)-quarks and a
heavier $s$-quark) the deconfinement transition is
 a rapid cross-over rather than a first order phase
transition with singularities in the bulk thermodynamic
observables \cite{Karsch:2004wd}. The critical temperature at $\mu_B = 0 $ for the rapid cross
over in the (2+1) flavor case was recently predicted to be $T_c= 172 \pm 11$~MeV \cite{Bernard:2004je}.

However, many QCD motivated calculations for low temperatures
and high baryon densities exhibit a strong first order phase
transition (with a phase coexistence region)
\cite{Alford:1997zt,Rapp:1997zu}.
These two limiting cases suggest that there exists a
critical  point (second order phase transition) \cite{Stephanov:1998dy} at the  end point of a
line of first order phase transitions.
Recently, the exploration of the phase diagram for large
temperatures and small, but non-vanishing, values of the
baryon chemical potential became possible through the
application of novel techniques, such as Ferrenberg-Swendsen
re-weighting \cite{Fodor:2001pe}, Taylor series expansions
\cite{Gavai:2003nn,Allton:2002zi} or simulations with an imaginary
chemical potential \cite{D'Elia:2002gd,Fodor:2002km}.
Although these techniques have allowed for considerable improvements,
the location of the critical point in the $T$ -- $\mu_B$ plane still has large
 theoretical uncertainties, due to the sensitivity to
 the quark masses and the lattice sizes used in the calculations.
 The predicted value of  $\mu_B^{endpoint}/ T_c$ varies between $\mu_B^{endpoint}/ T_c \simeq 1$ -- 3
~\cite{Gavai:2004sd,Fodor:2004nz,Ejiri:2003dc}, i.e.
$\mu_B^{endpoint} = $170 -- 420  MeV.

For the calculation presented in this work, 
we use a simple equation of state with the first order 
phase transition, which will allow us to compare our results to
previous hydrodynamic and hybrid calculations employing (1+1) 
dimensional \cite{BaDu00} and (2+1) dimensional \cite{TeLaSh01} hydrodynamic
models. 

Above the critical temperature ($T_c=160$ MeV at $\mu=0$ MeV), 
the  thermodynamical quantities are assumed to be 
determined by a QGP 
which is dominated by massless $u,d,s$ quarks and gluons. 
The pressure in QGP phase is given by 
\begin{eqnarray}
p_{\rm Q}&=&\frac{(32+21N_f)\pi^2}{180}T^4 \\ \nonumber 
	&&+ 
\frac{N_f}{2}\left ( \frac{\mu}{3}\right )^2T^2 + 
\frac{N_f}{4\pi^2}\left ( \frac{\mu}{3}\right )^4 - B, 
\label{Eq-p_qgp}
\end{eqnarray}
where $N_f$ is 3 and $B$ is the Bag 
constant \cite{SoHuKaRuPrVe97,HuSh98}.
For the hadronic phase we use a hadron gas equation
of state with excluded volume 
correction \cite{RiGoStGr91}. 
Here, the pressure for fermions is given by 
\begin{eqnarray}
p_{\rm H}^{\rm excl} (T,\{ \mu_i \} ) & = & \sum_i p^{\rm ideal}_i
(T,\mu_i- V_0 p^{\rm excl}(T,\{ \mu_i\})) \nonumber \\
  & = & \sum_i p^{\rm ideal}_i(T,\tilde{\mu}_i),
\label{Eq-p_had}
\end{eqnarray}
where $p^{\rm ideal}$ is the pressure of ideal hadron gas and $V_0$  
is excluded volume of hadrons whose radii are fixed to 0.7 fm.   
In the low-temperature region the well-established (strange and non-strange)
hadrons up to masses of $\sim2$~GeV are included in the EoS 
(see tables Tab.~\ref{bartab} and~\ref{mestab} for a detailed listing). 
Although heavy states are rare in
thermodynamical equilibrium, they have a larger entropy per particle
than light states, and therefore have considerable impact on the
evolution. In particular, hadronization is significantly faster
as compared to the case where the hadron gas consists of light mesons
only (see the discussion 
in~\cite{BSch,SoHuKaRuPrVe97,SoHuRu99,DumRi,feedback,CRS}).

\begin{table}
\begin{tabular}{cccccc}
\hline \hline
nucleon&delta&lambda&sigma&xi&omega\\  \hline \hline
$N_{938} $&$\Delta_{1232}$&$\Lambda_{1116}$&$\Sigma_{1192}$
&$\Xi_{1317}$&$\Omega_{1672}$\\
$N_{1440}$&$\Delta_{1600}$&$\Lambda_{1405}$&$\Sigma_{1385}$&$\Xi_{1530}$&\\
$N_{1520}$&$\Delta_{1620}$&$\Lambda_{1520}$&$\Sigma_{1660}$&$\Xi_{1690}$&\\
$N_{1535}$&$\Delta_{1700}$&$\Lambda_{1600}$&$\Sigma_{1670}$&$\Xi_{1820}$&\\
$N_{1650}$&$\Delta_{1900}$&$\Lambda_{1670}$&$\Sigma_{1775}$&$\Xi_{1950}$&\\
$N_{1675}$&$\Delta_{1905}$&$\Lambda_{1690}$&$\Sigma_{1790}$&$$&\\
$N_{1680}$&$\Delta_{1910}$&$\Lambda_{1800}$&$\Sigma_{1915}$&$$&\\
$N_{1700}$&$\Delta_{1920}$&$\Lambda_{1810}$&$\Sigma_{1940}$&$$&\\
$N_{1710}$&$\Delta_{1930}$&$\Lambda_{1820}$&$\Sigma_{2030}$&&\\
$N_{1720}$&$\Delta_{1950}$&$\Lambda_{1830}$&$$&&\\
$N_{1900}$&&$\Lambda_{2100}$&$$&&\\
$N_{1990}$&&$\Lambda_{2110}$&$$&& \\
$N_{2080}$ &&&&&\\
$N_{2190}$ &&&&&\\
$N_{2200}$ &&&&&\\
$N_{2250}$ &&&&&\\
\hline \hline
\end{tabular}
\caption{\label{bartab} Baryons and baryon-resonances treated in
the model. The corresponding
antibaryon states are included as well.}
\end{table}

\begin{table}
\begin{tabular}{cccccc}
\hline \hline 
$0^-$ & $1^-$ &$ 0^+$ &$ 1^+$ &$ 2^+$ & $(1^-)^*$\\ \hline \hline
 $\pi$ & $ \rho$ & $ a_0$ & $ a_1$ & $ a_2$ & $ \rho(1450)$ \\
 $K  $ &$   K^*$ & $ K_0^*$ & $ K_1^*$ & $ K_2^*$ &$ \rho(1700)$ \\
 $\eta$&$  \omega$& $ f_0 $&  $f_1$ & $ f_2 $ & $ \omega(1420)$ \\
 $\eta'$&  $\phi $&  $f_0^*$ & $ f_1'$& $ f_2'$ & $ \omega(1600)$ \\
\hline\hline
\end{tabular}
\caption{\label{mestab} Mesons and meson-resonances, sorted with
respect to spin and parity, treated in
the model.}
\end{table}

The hadronic states used in the EoS of our hydrodynamic calculation
are identical to those used in the microscopic model employed 
for the hadronic stage of the reaction (UrQMD,
see section~\ref{UrQMD_section}). This is necessary to ensure 
consistency between the properties of the hadron gas in the
hydrodynamic as well as in the microscopic picture and allows
for a smooth transition from one description to the other.
UrQMD additionally assumes a continuum of
color-singlet states called ``strings'' above the $m\simeq2$~GeV
threshold to model $2\rightarrow n$ processes and inelastic
processes at high CM-energy. For example, the annihilation of an
$\overline{p}$ on an $\Omega$ is described as excitation of two
strings with the same quantum numbers as the incoming hadrons,
respectively, which are subsequently mapped on known hadronic states
according to a fragmentation scheme. Since we shall be interested in
the dynamics of the $\Omega$-baryons emerging from the hadronization
of the QGP, it is unavoidable to treat string-formation.
The fact that string degrees of freedom are not taken into account
in the EoS~(\ref{Eq-p_qgp}) does not represent a problem in our case
since we focus on rapidly expanding systems where those degrees of
freedom can not equilibrate~\cite{belkbrand}.

The phase coexistence region 
is constructed employing Gibbs' conditions of phase equilibrium. The bag
parameter of $B=385$ MeV/fm$^3$ is chosen to yield the critical temperature
$T_C=160$~MeV at $\mu=0$.
In the coexistence region of QGP phase and hadron phase (i.e. the mixed phase) 
we introduce the fraction of the volume of the QGP phase, 
$\lambda(x_\mu)$ ($0 \leq \lambda \leq 1$) and parameterize energy 
density and baryon number density as 
\begin{eqnarray}
\epsilon_M(\lambda, T^*)& = & 
\lambda \epsilon_Q (T^*) + (1-\lambda)\epsilon_H(T^*), \nonumber\\ 
n_{BM}(\lambda, T^*) & = & \lambda  n_{BQ}(T^*) + 
(1-\lambda)n_{BH}(T^*), 
\end{eqnarray}
where $T^*=T^*(\mu)$ is the value of temperature on the phase boundary 
which is determined by Gibbs' conditions of phase equilibrium.
\begin{figure}[tbh]
\includegraphics[width=0.9\linewidth]{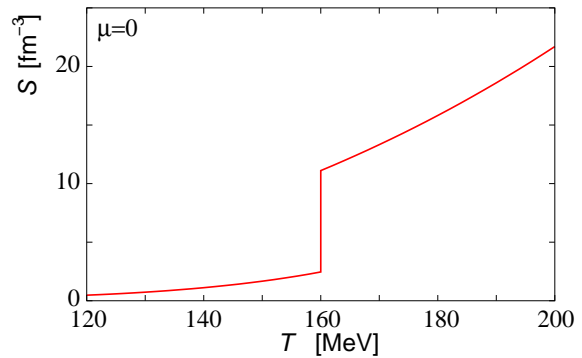}
\caption{Entropy density as a function of temperature at $\mu=0$.   
The critical temperature is 160 MeV.  
}
\label{Fig-eos_entro}
\end{figure}

In a forthcoming publication, 
we will discuss the EoS dependence of physical observables  
utilizing a realistic lQCD based equation of state, i.e.\ an EoS with a 
crossover phase transition 
at high $T$ and low $\mu$,  including the QCD critical point \cite{NoAs05}.    
We are currently in the process of constructing  such an EoS, which
will correctly capture critical phenomena   
around the QCD critical point \cite{NoAs05}.     
A simple parameterization of the EoS around the QCD critical point  
as presented in \cite{Brazil_QM} is unfortunately insufficient
for a description of these phenomena, even though it already provides
a marked improvement over currently used equations of state.

\subsection{Initial Conditions}
\label{sec-IC}
The initial conditions for the hydrodynamic calculation 
need to be determined either by adjusting an appropriate parametrization
to data or by utilizing other microscopic transport model predictions
for the early non-equilibrium phase of the heavy-ion reaction.

Numerous studies exist for finding the appropriate initial conditions 
for hydrodynamic models \cite{SoHuRu99,KoHeHuEsTu01,HiNa04,HiHeKhLaNa05,EsHoNiRuRa05}
-- usually such an initial condition is given by the parameterization of the
spatial distribution of the energy 
or entropy density and baryon number density at an initial time $\tau_0$. 
A comparison between final particle distributions calculated by the hydrodynamic 
model and experimental data can then be utilized to fix the values 
of the parameters for the initial conditions.  
However, this ansatz is problematic if no experimental data exist to tune
the initial conditions. Furthermore, one looses predictive and 
analytic power by treating the quantities governing the initial conditions as
free parameters.
Recently there have been several attempts to determine a set of initial conditions 
not from parameterizations and comparison to data, but via a calculation
using the color 
glass condensate (CGC) model for the initial state \cite{HiNa04,HiHeKhLaNa05} as
well as an approach combining perturbative QCD and the 
saturation picture \cite{EsHoNiRuRa05}. A study of elliptic flow by  
Hirano et al. has shown  that additional dissipation 
during the early QGP stage is required if an initial condition based on the 
CGC \cite{HiHeKhLaNa05} is used.
Another interesting fact which is found in hydrodynamic analyses at RHIC  
is that thermalization is achieved on very short timescales after the full overlap
of the colliding nuclei: none of the hydrodynamic calculations which have 
successfully addressed RHIC data at the top energy of $\sqrt{s_{NN}}=200$~GeV/nucleon
have initial times later than
$\tau_0 \sim 0.6$ fm \cite{HuKoHeRuVo01,HiTs02}.  
The physics processes leading to such a rapid thermalization have yet
to be unambiguously identified \cite{Mr05} -- note that hydrodynamics itself
cannot address the question of thermalization, since it relies on the assumption
of matter being in local thermal equilibrium.

For our calculation we use a simple initial condition which is parameterized based on 
a combination of wounded nucleon and binary collision scaling 
\cite{JaCo00,KoSoHe00,KoHeHuEsTu01}.  
Similar parameterizations have been used in various hydrodynamic models 
which have been successful in explaining  numerous
experimental observations at RHIC \cite{HuKoHeRuVo01,TeLaSh01,HiTs02}.  
We have chosen this common initial condition for our investigation in
order to describe the general features of our model and provide
a base-line comparison to previous calculations under similar assumptions --
in a subsequent publication we shall investigate the sensitivity of 
our results to particular variations and assumptions regarding the choice
of the initial conditions.

\begin{figure}[tbh]
\begin{minipage}[t]{80mm}
\includegraphics[width=0.9\linewidth]{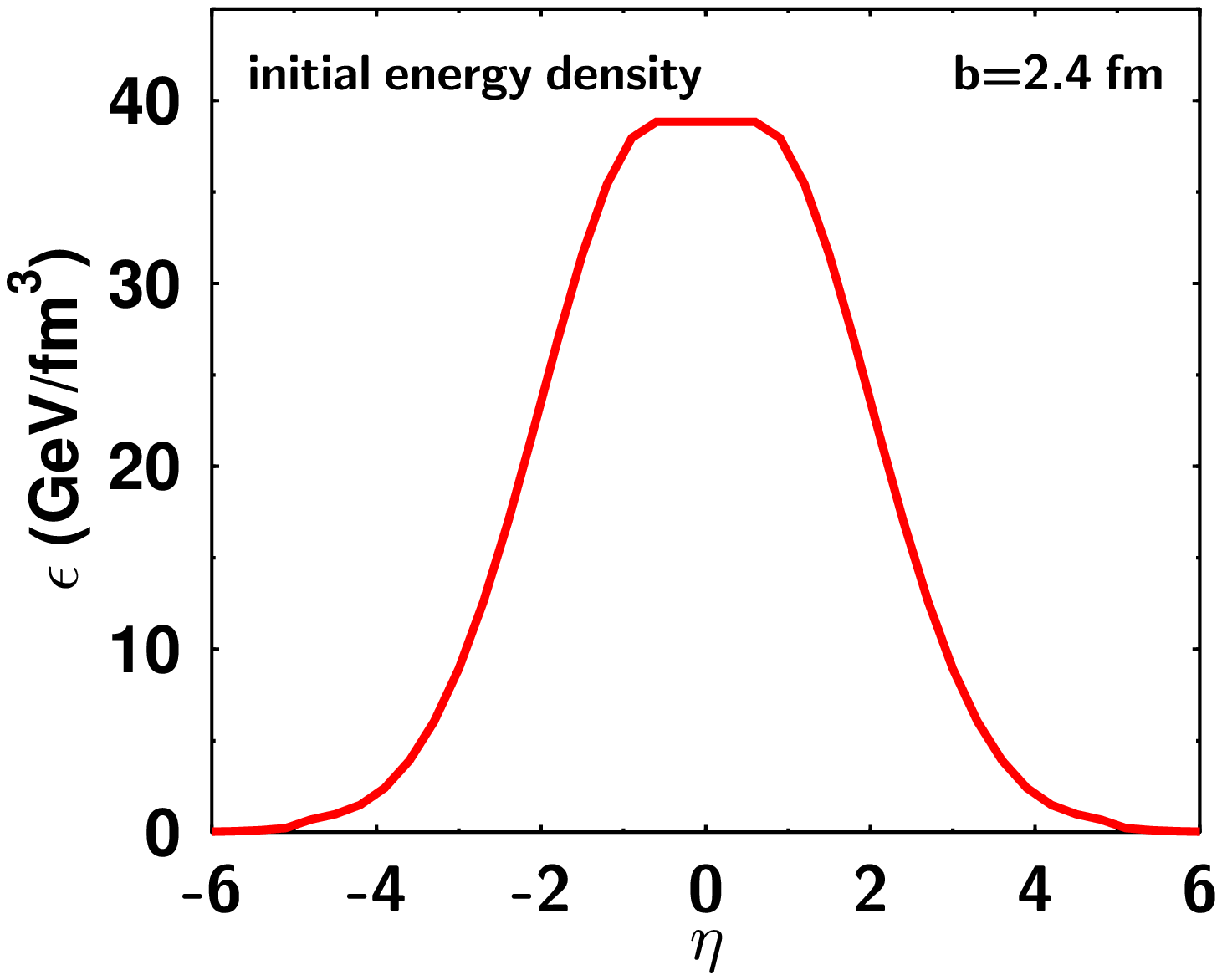}
\caption{Initial energy density in longitudinal direction at 
Au+Au $ \sqrt{s_{NN}}=200$ GeV central collisions (b=2.4 fm) in 
the case of Hydro + UrQMD. 
}
\label{Fig-ene_ini}
\end{minipage}
\hspace{3mm}
\begin{minipage}[t]{80mm}
\includegraphics[width=1.1\linewidth]{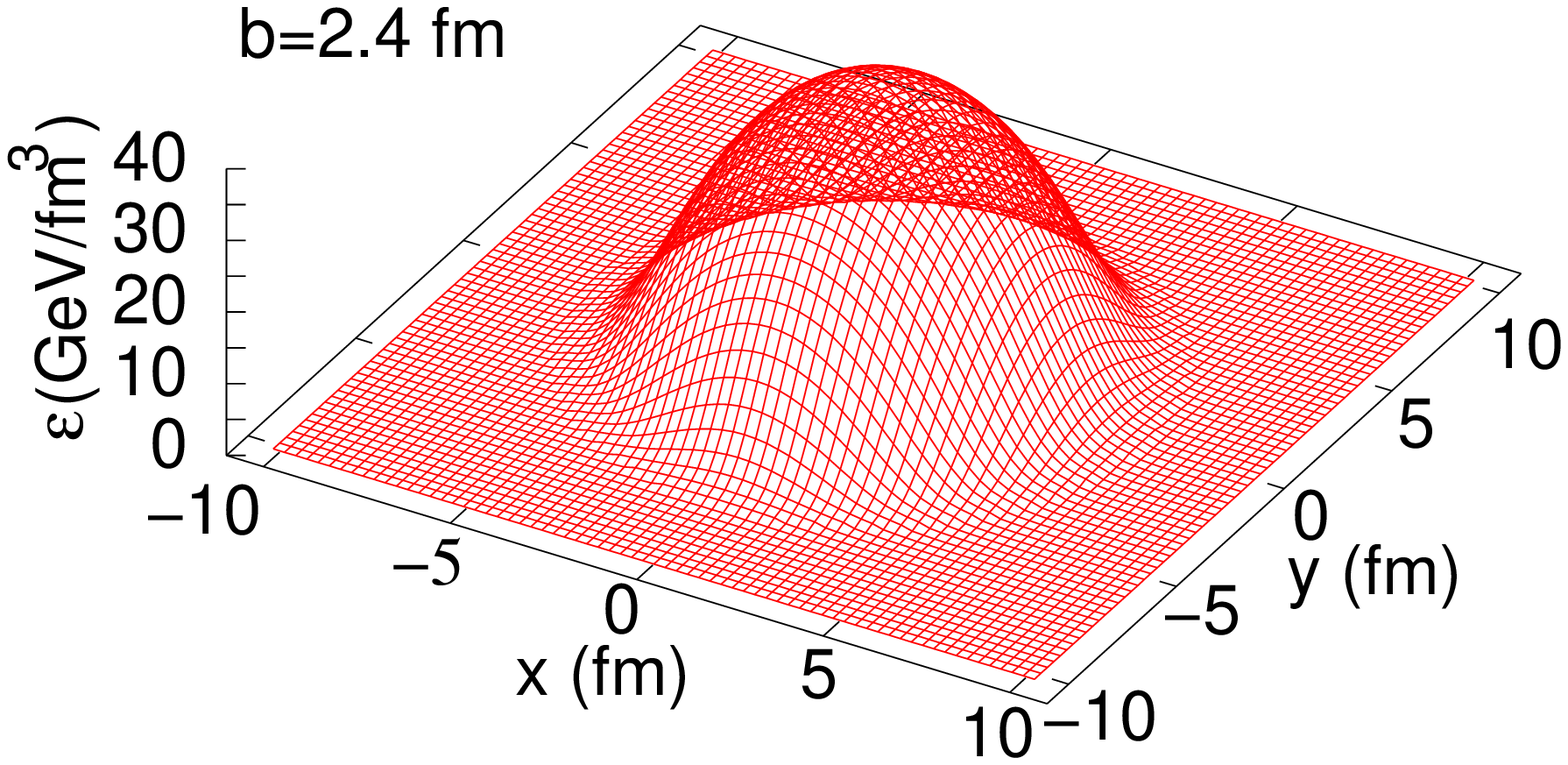}
\caption{Initial energy density on the transverse plane at   
Au+Au $ \sqrt{s_{NN}}=200$ GeV central collisions (b=2.4 fm) in 
the case of Hydro + UrQMD. 
}
\label{Fig-ene_ini_xy}
\end{minipage}
\end{figure}

We factorize the energy density and baryon number density 
distributions into longitudinal direction ($H(\eta)$) and 
the transverse plane ($W(x,y;b)$), which are given by  
\begin{eqnarray}
\epsilon(x,y,\eta)& = & \epsilon_{\rm max}W(x,y;b)H(\eta), \nonumber   \\
n_B(x,y,\eta)& = & n_{B{\rm max}}W(x,y;b)H(\eta), 
\end{eqnarray}
where $\epsilon_{\rm max}$ and $n_{B{\rm max}}$ are parameters which are 
maximum values of energy density and baryon number density.       
The longitudinal distribution is parameterized by  
\begin{equation}
H(\eta)=\exp \left [ -(|\eta|-\eta_0)/2 \sigma^2_\eta  \right ]
\theta (|\eta|-\eta_0), 
\end{equation}
where parameters $\eta_0$ and $\sigma_\eta$ are determined 
by comparison with experimental data of single particle 
distributions.
The function $W(x,y;b)$ on the transverse plane is determined by the 
superposition of wounded nucleon scaling which is characteristic of 
``soft" particle production processes and binary collision scaling 
which is characteristic of ``hard'' particle production processes \cite{HeKo02}. 
This function is normalized by $W(0,0;0)$.    
In the wounded nucleon scaling, the density of wounded nucleons is   
given by 
\begin{eqnarray}
\frac{d^2N_{\rm WN}}{ds^2}
&=& T_A(\vba) \cdot \left( 1 - e^{-T_B(\vbb) \sigma}\right) \nonumber \\
&+& T_B(\vbb)\cdot \left( 1 - e^{-T_A(\vba) \sigma}\right)\quad, 
\end{eqnarray}
where $\vba=\vs + b \cdot e_x$ ($\vbb=\vs-b \cdot e_x$), $\sigma$ is the 
total nucleon-nucleon cross section at Au+Au $\sqrt{s_{NN}}=200$ AGeV and 
set to 42 mb \cite{PHENIX03}.    
$T_A$ is the nuclear thickness function of nucleus $A$, 
\begin{equation}
T_A(\vs)=\int dz \rho_A(z, \vs), 
\end{equation}
where $\rho_A(z, \vs)$ is given by a Woods-Saxon parameterization of nuclear 
density,  
\begin{equation}
\rho_A(r) = \rho_0 \frac{1}{1 + e^{(r-R_A)/a}}.
\label{Eq-Woods-Saxon}
\end{equation}
In Eq.~(\ref{Eq-Woods-Saxon}) parameters $a$, $R_A$, $\rho_0$ are 
0.54 fm, 6.38 and 0.1688, respectively \cite{PHENIX03}.
On the other hand, the distribution of the number of binary 
collisions is given by  
\begin{equation}
\frac{d^2N_{\rm BC}}{ds^2} = \sigma \cdot T_A(\vba) T_B(\vbb). 
\end{equation}
Then total $W(x,y;b)=w\frac{d^2N_{\rm BC}}{ds^2} +(1-w) 
\frac{d^2N_{\rm WN}}{ds^2}$,      
where $w$ is the weight factor for binary scaling and is set to 0.6, again 
utilizing a comparison of experimental data of single particle spectra
to our model calculations.     
Figures \ref{Fig-ene_ini} and \ref{Fig-ene_ini_xy} show  
initial energy density in the longitudinal direction and on the transverse 
plane for Au+Au $\sqrt{s_{NN}}=200$ GeV central collisions 
for the case of a hybrid hydro+micro calculation.

As a starting point 
we set initial longitudinal flow to Bjorken's scaling solution \cite{Bjorken:1982qr} 
and neglect initial transverse flow.  
This is the simplest initial flow profile which will serve as basis 
for further investigation.   
For example, Kolb and Rapp discussed the possibility of existence of 
initial transverse flow  which improves the results for $P_T$ spectra and  
reduces the anisotropy in flow \cite{KoRa03}, however, at the expense of introducing an
additional parameter.
Utilizing a parameterized evolution model, it has been pointed out that a Landau-type 
initial longitudinal compression and re-expansion of matter 
is favorable for the description of Hanbury-Brown Twiss (HBT) 
correlation radii \cite{Th04}. 
This suggests that HBT analyses may be a sensitive tool for the determination
of the initial longitudinal flow distribution.     

In Tab. \ref{table:initial} parameters with which we reproduce 
single particle spectra at RHIC are listed. 
The parameters $\eta_0$ and $\sigma_\eta$ in longitudinal direction   
are determined mainly by the hadron rapidity distributions and 
do not strongly affect the transverse 
momentum distributions.
The parameters for the initial conditions need to be optimized separately
for the  pure hydro and hydro + UrQMD calculations.   
A comparison between the two sets
of initial conditions and possible physics implications can be found
in Sec. \ref{sec-results}.     
Comparing the initial energy density of our purely hydrodynamic
calculation to Ref.~\cite{KoRa03}  in which an 
initial condition was parameterized in terms of the
entropy- and baryon number density (weighted according to
the wounded nuclear model by 75\% and to the 
binary collision model by 25 \%) we find that our calculation
requires a significantly higher initial energy- or entropy density:  
the maximum value of the 
entropy density at Au + Au $\sqrt{s_{NN}}=200$ GeV in
central collision in~\cite{KoRa03} is 110 fm$^{-3}$, whereas 
our  value of the maximum entropy density is 176 fm$^{-3}$.
The difference between the two values is most likely 
due to the treatment of resonance decays -- these 
are explicitly treated in~\cite{KoRa03},
but are neglected in our case. Note that we consider our 
purely hydrodynamic calculation solely as a baseline against
which we can compare the full hydro+micro model, which of course
contains a proper treatment of resonance decays.
A third hydrodynamic model implementation described in~\cite{HiNa04} 
uses an initial condition purely based on binary collisions. 
The maximum value of energy density in that calculation (which
also treats resonance decays)
is found to be $\sim 45$ GeV/fm$^3$, similar to the value we
obtain for our hydro+micro calculation.

\begin{table}[tbh]
\begin{tabular}{|c|c|c|c|c|c|}
\hline \hline
  & $\tau_0$   &   $\epsilon_{\rm max}$  &
     $n_{B {\rm max}}$  & $\eta_0$ & $\sigma_\eta$  \\ 
  &  (fm/$c$)      &   (GeV/fm$^3$) &
      (fm$^{-3}$) &  &   \\ \hline

     pure hydro & 0.6    & 55   & 0.15   & 0.5   & 1.5  \\ \hline
      hydro + UrQMD & 0.6   & 40    &  0.15  & 0.5   & 1.5  \\ \hline
\end{tabular}
\caption{Parameters for initial conditions of pure hydro and hybrid model.}
\label{table:initial}
\end{table}

\subsection{Hadronization and the transition to microscopic dynamics}
Having specified the initial conditions on the $\tau=\tau_i$ hypersurface
and the
EoS, the hydrodynamical solution in the forward light-cone is determined
uniquely.  
We assume that a freezeout process happens when a temperature in a 
volume element of fluid is equal to a freezeout temperature $T_f$ 
in the pure 3-D hydrodynamic model.     
In a hybrid model the transition from 
macroscopic to microscopic dynamics takes place at a switching 
temperature $T_{\rm SW}$. 
The feezeout and switching temperatures respectively can be
treated as parameters and determined  
by comparison to experimental data on single particle spectra. 

Due to our use of  Lagrangian hydrodynamics, 
grid points move along the flux of the fluid and  
are not represented by a fixed coordinate-space lattice any more.  
Therefore it is non-trivial to estimate of the number of 
particles flowing through the freezeout hypersurface.   
Here we start from a simple case \cite{BlOl90}:    
suppose that the number of particles $N(\tau)$  exists in the   
enclosed volume $\Omega$ that is bounded by a closed surface 
$S(\tau)$ at time $\tau$. $N(\tau)$ is given by  
\begin{equation}
N(\tau)=\int_{\Omega(\tau)} d^3r n(\vecr,\tau), 
\end{equation}
where $n(\vecr, \tau)$ is particle number density.
At time $\tau + \delta \tau$, the number of particle $N(\tau)$ 
changes to  
\begin{equation}
\frac{dN}{d \tau} = \int_{\Omega(\tau)} d^3r \frac{\partial n}{\partial \tau} 
+ \frac{1}{d \tau} \int_{\delta \Omega}d^3r n(\vecr,\tau), 
\label{Eq-num1}
\end{equation}
where the volume varies to $\Omega + \delta \Omega$.      
Utilizing current conservation, $\frac{\partial n}{\partial \tau} + \nabla \cdot \vj$ 
($\vj$ is a current of particle),    
Eq.(\ref{Eq-num1}) is rewritten as  
\begin{equation}
\frac{dN}{d\tau}=-\int_{S(\tau)}d^2s \vj \cdot \vn+ \int _{S(\tau)} d^2s 
\frac{d\zeta}{d \tau}n,    
\label{Eq-num2}
\end{equation}
where $\vn$ is the normal vector of surface element $d^2s$ and 
$d\zeta$ is the distance between the surface of $\Omega(\tau)$  
and that of $\Omega(\tau)+\delta \Omega$.  
In Eq.\ (\ref{Eq-num2}), $dN/d\tau$ is the number of particles 
which cross 
the surface $S(\tau)$ during $d\tau$.  
Then total number of particles through the hypersurface $\Sigma$ 
which is the set of surfaces $\{S(\tau)\}$,  
\begin{equation}
N=\int _\Sigma j^\mu d\sigma_\mu, 
\label{Eq-num3}
\end{equation}
where $j^0=n$, $d\sigma_0=d^3r$, $d \vsigma =d\tau \cdot d^2s \vn$.
If we write
\begin{equation}
j^\mu=\frac{d^3P}{E}\frac{g_h}{(2\pi)^3}
\frac{1}{\exp[(P_\nu u^\nu-\mu_f)/T_f]\pm 1} P^\mu
\end{equation}
for the current $j^\mu$ in Eq.\ (\ref{Eq-num3}), we obtain the   
Cooper-Frye formula \cite{CF}  
\begin{eqnarray}
&& E\frac{dN}{d^3P} = \nonumber \\
&& \sum_h\frac{g_h}{(2\pi)^3} \int_\Sigma d\sigma_\mu
P^\mu \frac{1}{\exp[(P_\nu u^\nu-\mu_f)/T_f]\pm 1}, 
\label{Eq-CF}
\end{eqnarray}
where $g_h$ is a degeneracy factor of hadrons and $T_f$ and $\mu_f$ are 
the freezeout temperature and chemical potential.  
In other words we obtain $d\sigma_\mu$ by estimating 
the normal vector on the freezeout hypersurface $\Sigma$. 
Using Eq.\ (\ref{Eq-CF}), we then can
 calculate all particle distributions.

In order to pass on the distribution~(\ref{Eq-CF}) in the macroscopic model 
to the microscopic model, 
we first calculate the multiplicities $N_i$ for each particle species $i$,  
by integration of 
Eq.\ (\ref{Eq-CF}) in which $T_f$ and $\mu_f$ are changed to the
switching temperature $T_{\rm SW}$ 
and chemical potential $\mu_{\rm SW}$ over   
space-time ($\tau$, \vecr, $\eta$) and momentum space ($\vPT$, $y$).
$N_i$ is rounded to an integer value since the hadronic transport 
model described in the next section deals with real particle degrees
of freedom.
The distribution~(\ref{Eq-CF}) divided by $N_i$ is used as
probability distribution to randomly generate the space-time and
momentum-space coordinates for $N_i$ hadrons of species $i$.
This distribution serves as an input for the hadronic transport model 
for a single event. The sampling procedure can be repeated to generate
a sequence of events as starting points for the microscopic calculation.    
Each event-sampling  produces a different set of space-time and momentum coordinates 
as input of the microscopic model, however,  the total multiplicity for each species $i$
remains constant for a given hydrodynamic switching hyper-surface.    
In a more realistic calculation, particle number fluctuations
should be taken into account as well.   
The reverse process, absorbing microscopic particles into the hydrodynamic medium
is being neglected -- it has been shown in \cite{Teaney:2001av} that for rapidly expanding
systems these contributions are negligible.

\subsection{Microscopic dynamics: the UrQMD approach}
\label{UrQMD_section}
The ensemble of hadrons generated accordingly is
then used as initial condition for the microscopic transport model
Ultra-relativistic Quantum Molecular Dynamics (UrQMD) 
\cite{uqmdref1}. The UrQMD approach is closely related to hadronic 
cascade \cite{cascade}, Vlasov--Uehling--Uhlenbeck \cite{vuu}
and (R)QMD transport models \cite{qmd}.
We shall describe here only the part of the model that is important
for the application at hand, namely the evolution of an expanding hadron gas
in local equilibrium at a temperature of about $T_C\sim160$~MeV.
The treatment of high-energy hadron-hadron scatterings, as it occurs in
the initial stage of ultrarelativistic collisions, is not discussed here.
A complete description of the model and detailed comparisons to experimental
data can be found in~\cite{uqmdref1}.

The basic degrees of freedom are hadrons modeled as Gaussian wave-packets,
and strings, which are used to model the fragmentation of high-mass
hadronic states via the Lund scheme \cite{lund}.
The system evolves as a sequence of binary collisions or $2-N$-body decays 
of mesons, baryons, and strings.

The real part of the nucleon optical potential, i.e.\ a mean-field,
can in principle be included in UrQMD for the dynamics of baryons 
(using a Skyrme-type interaction with a hard equation of state). However,
currently no mean field for mesons (the most abundant hadrons in our
investigation) are implemented.
Therefore, we have not accounted for mean-fields in the equation
of motion of the hadrons. 
To remain consistent, mean fields were also not taken into account
in the EoS on the fluid-dynamical side. Otherwise, pressure equality (at given
energy and baryon density) would be destroyed.
We do not expect large modifications of the results presented here due to
the effects of mean fields, since the ``fluid'' is not very dense after 
hadronization and current experiments at SIS and AGS only point
to strong medium-dependent properties of mesons (kaons in particular)
for relatively low incident beam energies 
($E_{lab} \le 4$~GeV/nucleon) \cite{gsikaonen}. 
Nevertheless, mean fields will have to be included in the future.
 
Binary collisions are performed in a point-particle sense:
Two particles collide if their minimum distance $d$, 
i.e.\ the minimum relative 
distance of the centroids of the Gaussians during their motion, 
in their CM frame fulfills the requirement: 
\begin{equation}
 d \le d_0 = \sqrt{ \frac { \sigma_{\rm tot} } {\pi}  }  , \qquad
 \sigma_{\rm tot} = \sigma(\sqrt{s},\hbox{ type} ).
\end{equation}
The cross section is assumed to be the free cross section of the
regarded collision type ($N-N$, $N-\Delta$, $\pi-N$ \ldots).

The UrQMD collision term contains 53 different baryon species
(including nucleon, delta and hyperon resonances with masses up to 2 GeV) 
and 24 different meson species (including strange meson resonances), which
are supplemented by their corresponding anti-particle 
and all isospin-projected states.
The baryons and baryon-resonances which can be populated in UrQMD are listed
in table~\ref{bartab}, the respective mesons in table~\ref{mestab} -- 
full baryon/antibaryon symmetry is included (not shown in the table), both,
with respect to the included hadronic states, as well as with respect to
the reaction cross sections.
All hadronic states can be produced in string decays, s-channel
collisions or resonance decays. 

Tabulated and parameterized experimental 
cross sections are used when available. Resonance absorption, decays 
and scattering are handled via the principle of detailed balance. 
If no experimental information is
available, the cross section is either  calculated via
an One-Boson-Exchange (OBE) model or via a modified additive quark model
which takes basic phase space properties into account.

In the baryon-baryon sector, the total and elastic proton-proton and 
proton-neutron cross sections are well known \cite{PDG96}. 
Since their functional dependence on $\sqrt{s_{NN}}$ shows
a complicated shape at low energies, UrQMD uses a table-lookup for those
cross sections. However, many cross sections involving strange baryons and/or 
resonances are not well known or even experimentally accessible -- for these
cross sections the additive quark model is widely used.

As we shall see later, the most important reaction channels 
in our investigation are meson-meson and
meson-baryon elastic scattering and resonance formation. 
For example, the total meson-baryon cross section for
non-strange particles is  given by
\begin{eqnarray}
\label{mbbreitwig}
\sigma^{MB}_{tot}(\sqrt{s}) &=& \sum\limits_{R=\Delta,N^*}
       \langle j_B, m_B, j_{M}, m_{M} \| J_R, M_R \rangle \nonumber \\
&&\times        \frac{2 S_R +1}{(2 S_B +1) (2 S_{M} +1 )} \nonumber \\
&&\times        \frac{\pi}{p^2_{CMS}}\, 
        \frac{\Gamma_{R \rightarrow MB} \Gamma_{tot}}
             {(M_R - \sqrt{s})^2 + \frac{\Gamma_{tot}^2}{4}}
\end{eqnarray}
with the total and partial $\sqrt{s}$-dependent decay widths $\Gamma_{tot}$ and
$\Gamma_{R \rightarrow MB}$. 
The full decay width $\Gamma_{tot}(M)$ of a resonance is 
defined as the sum of all partial decay widths and depends on the
mass of the excited resonance:
\begin{equation}
\Gamma_{tot}(M)  
\label{gammatot}
       \,=\, \sum  \limits_{br= \{i,j\}}^{N_{br}} \Gamma_{i,j}(M) \quad.
\end{equation}
The partial decay widths $\Gamma_{i,j}(M)$ for the decay into the 
final state with particles $i$ and $j$ is given by
\begin{eqnarray}
\label{gammapart}
\Gamma_{i,j}(M)
        &=&
       \Gamma^{i,j}_{R} \frac{M_{R}}{M}
        \left( \frac{\langle p_{i,j}(M) \rangle}
                    {\langle p_{i,j}(M_{R}) \rangle} \right)^{2l+1}\nonumber \\
&\times&         \frac{1.2}{1+ 0.2 
        \left( \frac{\langle p_{i,j}(M) \rangle}
                    {\langle p_{i,j}(M_{R}) \rangle} \right)^{2l} } \quad,
\end{eqnarray}
here $M_R$ denotes the pole mass of the resonance, $\Gamma^{i,j}_{R}$
its partial decay width into the channel $i$ and $j$ at the pole and
$l$ the decay angular momentum of the final state.
All pole masses and partial decay widths at the pole are taken from the Review
of Particle Properties \cite{PDG96}. 
$\Gamma_{i,j}(M)$ is constructed in such a way that 
$\Gamma_{i,j}(M_R)=\Gamma^{i,j}_R$ is fulfilled at the pole.
In many cases only crude estimates for $\Gamma^{i,j}_R$ are given
in \cite{PDG96} -- the partial decay widths must then be fixed by
studying exclusive particle production in elementary proton-proton
and pion-proton reactions. 
Therefore, e.g., the total pion-nucleon cross section depends on the
pole masses, widths and branching ratios of all $N^*$ and $\Delta^*$
resonances listed in table~\ref{bartab}. 
Resonant meson-meson scattering
(e.g. $\pi + \pi \to \rho$ or $\pi + K \to K^*$)
is treated in the same formalism.

In order to correctly treat equilibrated matter~\cite{belkbrand}
(we repeat that the hadronic
matter with which UrQMD is being initialized in our approach is in local
chemical and thermal equilibrium), the principle of detailed balance is
of great importance.
Detailed balance is based on time-reversal invariance 
 of the matrix element of the reaction. It is most commonly found
in textbooks in the form:
\begin{equation}
\label{dbgl3}
\sigma_{f \rightarrow i } \,=\, \frac{\vec{p}_i^2}{\vec{p}_f^2} \frac{g_i}{g_f}
\sigma_{i \rightarrow f} \quad ,
\end{equation}
with $g$ denoting the spin-isospin degeneracy factors.
UrQMD applies the general principle of detailed balance to the 
following two process classes:
\begin{enumerate}
\item 
Resonant meson-meson and meson-baryon interactions: Each resonance created
via a meson-baryon or a meson-meson annihilation may again decay into
the two hadron species which originally formed it. This symmetry is only
violated in the case of three- or four-body decays and string fragmentations, 
since N-body collisions with (N$>2$) are not implemented in UrQMD. 
\item
Resonance-nucleon or resonance-resonance interactions: the excitation
of baryon-resonances in UrQMD is handled via parameterized cross sections
which have been fitted to data. The reverse reactions usually have not
been measured - here the principle of detailed balance is applied.
Inelastic baryon-resonance de-excitation is the only method in UrQMD
to absorb mesons (which are {\em bound} in the resonance). Therefore
the application of the detailed balance principle is of crucial
importance for heavy nucleus-nucleus collisions.
\end{enumerate}

Equation~(\ref{dbgl3}), however, is only valid in the case of stable
particles with well-defined masses. Since in UrQMD detailed balance
is applied to reactions involving resonances with finite lifetimes
and broad mass distributions, equation~(\ref{dbgl3}) has to be 
modified accordingly. For the case of one incoming resonance the
respective modified detailed balance relation has been derived
in \cite{danielewicz91a}. Here, we generalize this expression for
up to two resonances in both, the incoming and the outgoing channels.

The differential cross section for the reaction 
$(1\,,\,2) \rightarrow (3\,,\,4)$ is given by:
\begin{equation}
\label{diffcx1}
{\rm d} \sigma_{12}^{34} \,=\,
    \frac{| {\cal M} |^2}{64 \pi^2 s} \, \frac{p_{34}}{p_{12}} 
        \,{\rm d \Omega}\,
     \prod_{i=3}^4    \delta(p_i^2 -M_i^2) {\rm d}p_i^2  \quad,
\end{equation}
here the $p_i$ in the $\delta$-function denote four-momenta.
The $\delta$-function ensures that the particles are on mass-shell,
i.e. their masses are well-defined. If the particle, however, has 
a broad mass distribution, then the $\delta$-function
must be substituted by the respective mass distribution (including
an integration over the mass):
\begin{equation}
\label{diffcx2}
{\rm d} \sigma_{12}^{34} \,=\,
    \frac{| {\cal M} |^2}{64 \pi^2 s} \, \frac{1}{p_{12}} 
        \,{\rm d \Omega}\,
     \prod_{i=3}^4  p_{34} \cdot
   \frac{ \Gamma}{\left(m-M_i\right)^2+ \Gamma^2/4} \frac{{\rm d} m}{2\pi}
\, .
\end{equation}
Incorporating these modifications into equation~(\ref{dbgl3}) and
neglecting a possible mass-dependence of the matrix element we
obtain:
\begin{eqnarray}
\label{uqmddetbal}
     \frac{ {\rm d} \sigma_{34}^{12} }{{\rm d} \Omega }   
      & =& \frac{\langle p_{12 }^2 \rangle   }
       {\langle p_{34 }^2 \rangle  } \,
       \frac{(2 S_1 + 1) (2 S_1 + 1)}
            {(2 S_3 + 1) (2 S_4 + 1)}\nonumber \\
&\times&      \sum_{J=J_-}^{J_+} 
      \langle j_1 m_1 j_2 m_2 \| J M \rangle \, 
        \frac{ {\rm d} \sigma_{12}^{34} }{{\rm d} \Omega } \, .
\end{eqnarray}
Here, $S_i$ indicates the spin of particle $i$ and 
the summation of the Clebsch-Gordan-coefficients is over the isospin of the
outgoing channel only. For the incoming channel, isospin is 
treated explicitly. The summation limits are given by:
\begin{eqnarray}
  J_- &=& \max \left( |j_1-j_2|,  |j_3-j_4| \right) \\  
  J_+ &=& \min \left( j_1+j_2,  j_3+j_4     \right)  \quad.
\end{eqnarray}
The integration over the mass distributions of the resonances  
in equation~(\ref{uqmddetbal}) has been denoted by the brackets 
$\langle\rangle$,
e.g.
\begin{eqnarray*}
&& p_{3,4}^2 \Rightarrow \langle p_{3,4}^2 \rangle \\
&& = \,
 \int  \int 
  p_{CMS}^2(\sqrt{s},m_3,m_4)\, A_3(m_3) \, A_4(m_4) \, 
  {\rm d} m_3\; {\rm d} m_4 
\end{eqnarray*}
with the  
mass distribution $A_r(m)$ given by a free Breit-Wigner distribution
with a mass-dependent width according to equation~(\ref{gammatot}):
\begin{equation}
\label{bwnorm}
A_r(m) \, = \, \frac{1}{N} 
        \frac{\Gamma(m)}{(m_r - m)^2 + \Gamma(m)^2/4}
\end{equation}
with
\begin{displaymath}
        \lim_{\Gamma \rightarrow 0} A_r(m) = \delta(m_r - m) \, ,
\end{displaymath}
and the normalization constant
\begin{equation}
\label{bwnorm_norm}
N \,=\, \int\limits_{-\infty}^{\infty} 
\frac{\Gamma(m)}{(m_r - m)^2 + \Gamma(m)^2/4} \, {\rm d}m \, .
\end{equation}
Alternatively one can also choose a Breit-Wigner distribution with a fixed
width, the normalization constant then has the value $N=2\pi$.

The most frequent applications of equation~(\ref{uqmddetbal}) in UrQMD
are the processes $\Delta_{1232}\, N \to N\, N$ and
$\Delta_{1232}\, \Delta_{1232} \to N\, N$.

\section{Results}
\label{sec-results}
\subsection{Modeling of the entire reaction dynamics in three-dimensional
hydrodynamics}

The purpose of this section is to establish a baseline against which we
can later compare the results of our hybrid macro+micro model. In addition
we wish to demonstrate that our novel implementation of the 3+1 dimensional
relativistic hydrodynamic model, utilizing a Lagrangian grid, is capable
of reproducing the results of previous hydrodynamic calculations. Many
of these previous calculations found in the literature have focused on 
one or two particular observables -- here we wish to conduct a consistent
analysis of all relevant single-particle distributions which can be
addressed by a hydrodynamic calculation.

In order to obtain qualitative results which are comparable to
experimental data, we have to determine the parameters of the initial 
conditions. 
This is accomplished by adjusting the parameters to fit
the single particle spectra for the most central Au+Au collisions 
at $\sqrt{s_{NN}}=200$~GeV.

\begin{figure}[tbh]
\includegraphics[width=0.9\linewidth]{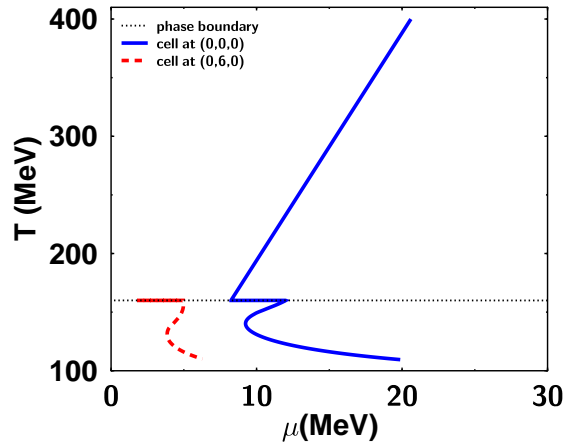}
\caption{The isentropic trajectories on $T-\mu$ plane.  Solid (Dashed) line 
stands for the behavior of sell which is located at 
$(x,y,\eta)=(0,0,0)(=(0,6,0))$  at initial time. Dotted line represents 
phase boundary.}
\label{Fig-hydro_tmu}
\end{figure}

First we show the behavior of isentropic trajectories in the $T-\mu$ plane  
for Au+Au $\sqrt{s_{NN}}=200$ GeV central collisions in Fig.~\ref{Fig-hydro_tmu}. 
The dotted line stands for the phase boundary between the QGP and the hadronic phase
(Note that due to small baryochemical potentials, 
the phase boundary is an almost flat line at $T_C=160$ MeV).  
Apart from the central cell, we also investigate the isentropic trajectory
of a cell close to the surface of the initially produced QGP.
Whereas the isentropic trajectory of the central cell located at 
$(0,0,0)$ starts in the QGP phase (solid line), 
the cell at the initial surface of the QGP (dashed line) only exhibits
an evolution from the mixed phase to the hadronic phase. 
Both trajectories are terminated at freezeout temperature, 
$T_{\rm f}=110$ MeV.  

\begin{figure}[tbh]
\includegraphics[width=0.9\linewidth]{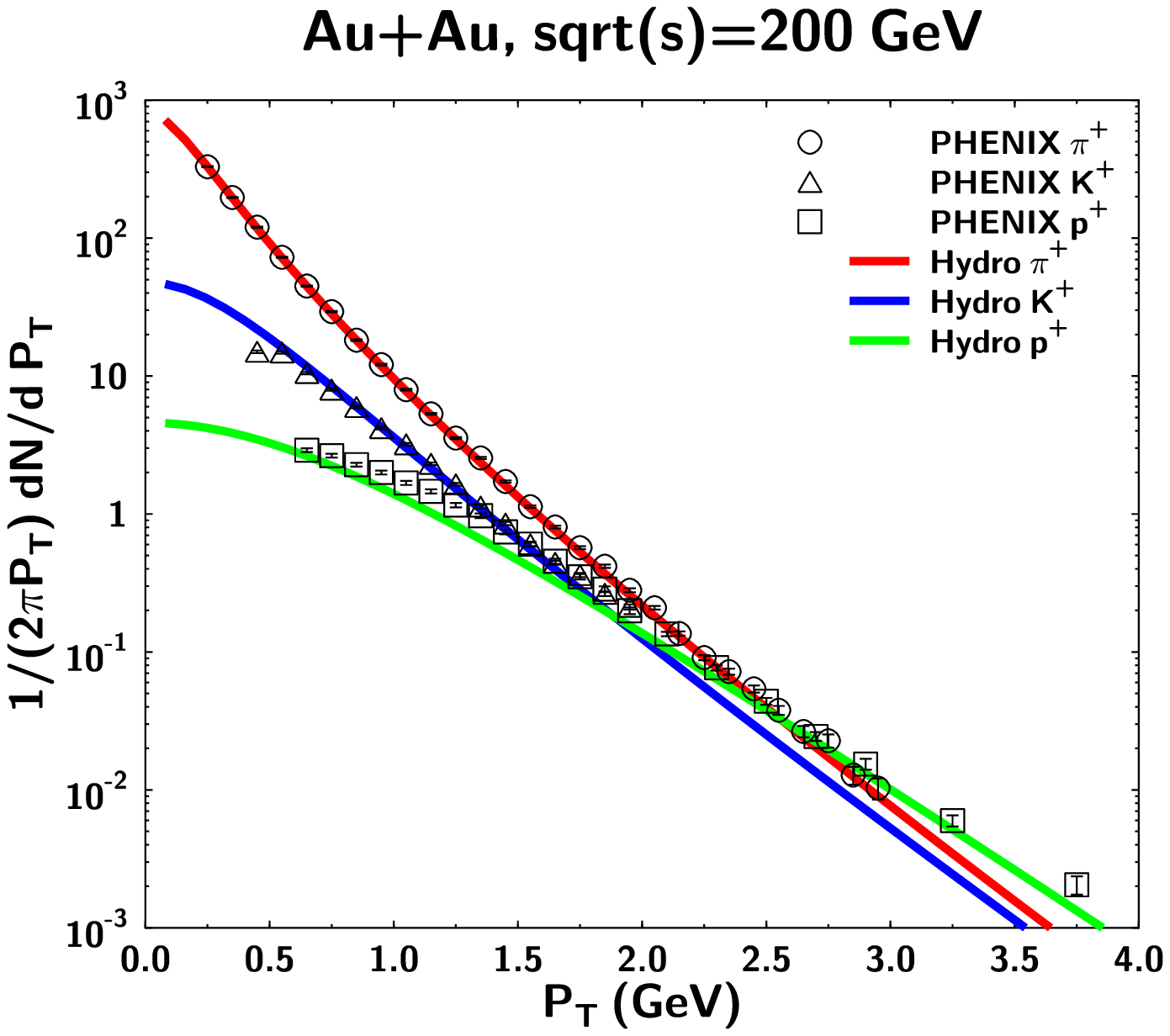}
\caption{$P_T$ spectra for $\pi^+$, $K^+$, $p$ at central collisions 
with PHENIX data \cite{PHENIX_PT}. For proton we rescale our result 
using the ration at chemical temperature (See the text).
}
\label{Fig-hydro_pt}
\includegraphics[width=0.9\linewidth]{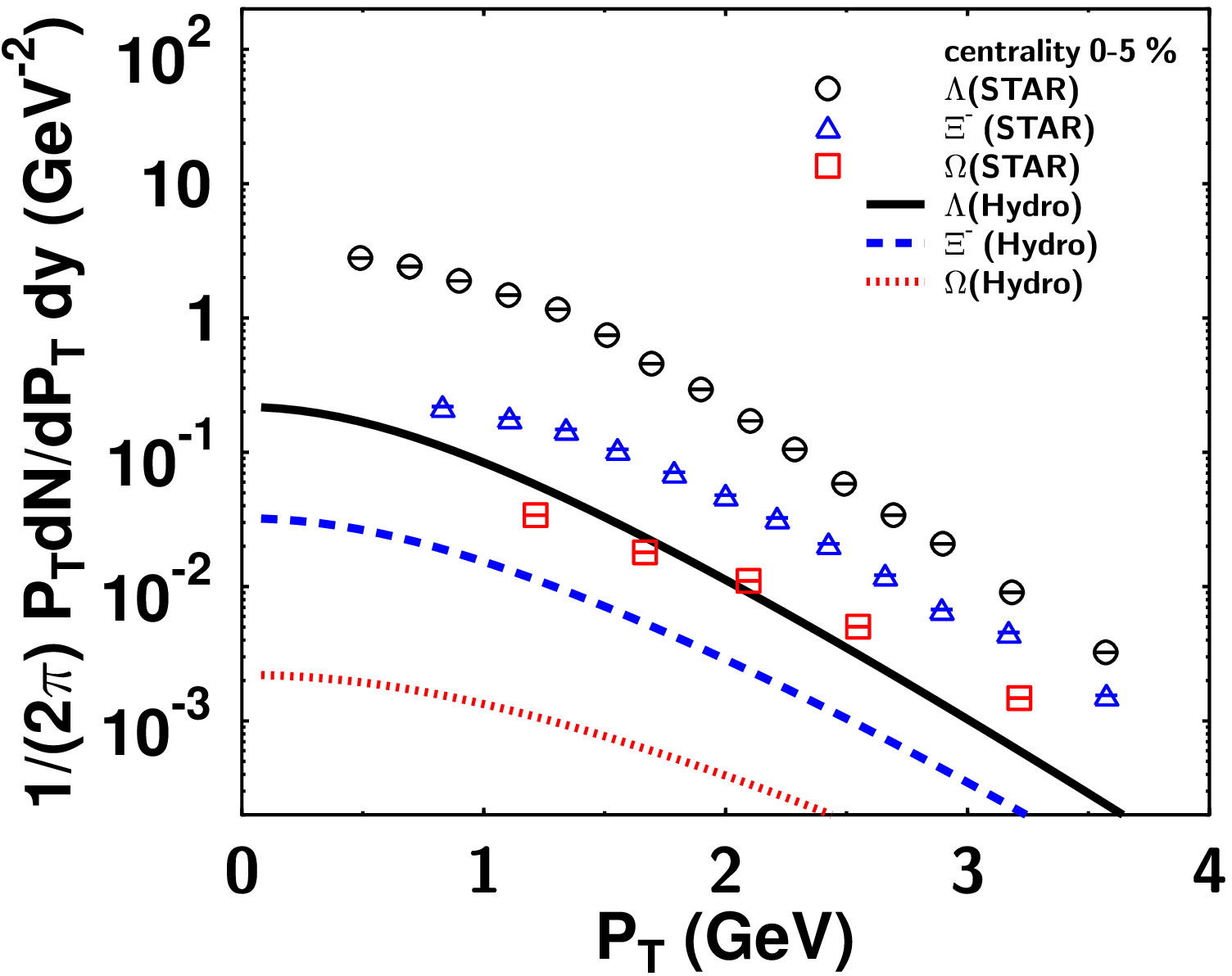}
\caption{$P_T$ spectra for multistrange baryons at central collisions 
with STAR data \cite{STAR_strange1}. In this calculation (pure hydro) 
the additional procedure for normalization is not done.  
}
\label{Fig-hydro_pt_ms}
\end{figure}

Figure \ref{Fig-hydro_pt} shows the 
$P_T$ spectra of $\pi$, $K$ and $p$ in Au + Au at $\sqrt{s_{NN}}=200$ GeV 
for central collisions. 
Our calculation succeeds in  reproducing  the $\pi$ spectra measured    
by the PHENIX collaboration \cite{PHENIX_PT} up to $P_T \sim 2$ GeV. 
However, due to the model assumption of chemical 
equilibrium up to the (low) kinetic freezeout temperature, 
we fail to obtain the correct normalization and hadron ratios, 
even though the shape of the $P_T$ spectra of $p$ and multistrange 
baryons (shown in Fig.~\ref{Fig-hydro_pt_ms})  is close to experimental data. 
In order to obtain the proper normalization for the $p$
spectra and match the 
experimentally measured hadron ratios
we adopt a procedure outlined in Ref.~\cite{hadron-ratio},
which renormalizes the $P_T$ spectra using the $p$ to $\pi$ ratio 
at the critical temperature to fix the normalization of the proton
spectra. It is straightforward to extend this procedure to hyperons
and multi-strange baryons as well, even though we choose to show
the real, unrenormalized, result for the multi-strange baryons
in Fig. \ref{Fig-hydro_pt_ms} to elucidate the situation prior to
renormalization.

The need for renormalizing the $p$ spectra 
suggests that the assumption of a continuous chemical 
equilibrium until kinetic freezeout  
is not realistic and that an improved treatment of the freeze-out process 
is required. 
One method to deal with the separation of chemical and
thermal freeze-out is the {\em partial chemical equilibrium model (PCE)}
\cite{HiTs02,Te02,KoRa03}:
below a chemical freeze-out temperature $T_{ch}$ one
introduces a chemical potential for each hadron whose
yield is supposed to be frozen-out at that temperature.
The PCE approach can account for the proper normalization
of the spectra, however, it fails to reproduce the
transverse momentum and mass dependence of the elliptic 
flow \cite{PHENIX_white}.  
In section~\ref{subsec-hybrid}, 
we shall utilize our hybrid hydro+micro model 
to decouple chemical and kinetic freeze-out. In these
hybrid approaches
\cite{BaDu00,TeLaSh01,HiHeKhLaNa05} 
the freeze-out occurs
sequentially as a result of the microscopic evolution and 
flavor degrees of freedom are treated explicitly through the
hadronic cross sections of the microscopic transport, leading
to the proper normalization of all hadron spectra.

\begin{figure}[tbh]
\includegraphics[width=0.9\linewidth]{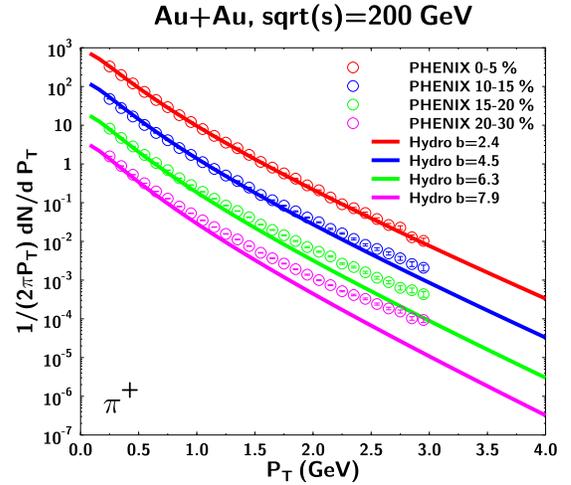}
\caption{Centrality dependence of $P_T$ spectra of $\pi^+$ with 
PHENIX data \cite{PHENIX_PT}.  The $P_T$ spectra at 10--15 \%, 15--20 
\% and  20--30 \% are divided by 5, 25 and 200, respectively.
}
\label{Fig-hydro_pt_cent}
\end{figure}

In Fig.~\ref{Fig-hydro_pt_cent} the centrality dependence of 
$P_T$ spectra for $\pi^+$ is shown. 
The impact parameters are set to $b=$2.4, 4.5, 6.3, 7.9 fm corresponding
to 0-6, 10-15, 15-20, 20-30 \% centrality, respectively.
These values are estimated via the number of 
nucleon-nucleon binary collisions and the number of participant nucleons 
in Ref.~\cite{PHENIX_PT}.      
The centrality dependence is determined simply by the collision 
geometry -- no additional parameter is necessary for our finite 
impact parameter collision calculations in  Sec.~\ref{sec-IC}.   
Our results are 
consistent with the experimental data \cite{PHENIX_PT} in the 
low transverse momentum region ($0<P_T<1$ GeV) for all 
centralities.  
We observe that in peripheral collisions the difference between 
experimental data and our calculations  appears at lower $P_T$ 
compared to central collisions.
This difference is indicative of the diminished importance of the soft, 
collective, physics described by hydrodynamics 
compared to the contribution of jet-physics in peripheral 
events. 

\begin{figure}[thb]
\includegraphics[width=0.9\linewidth]{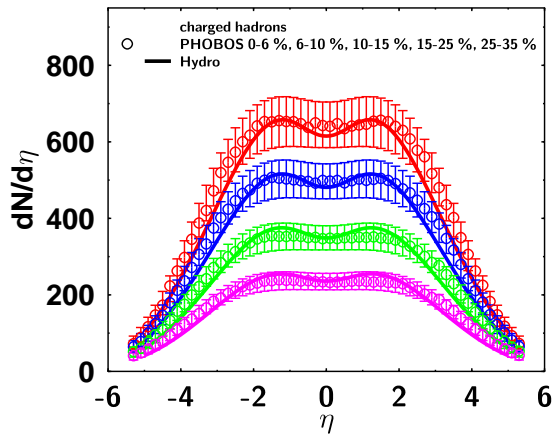}
\caption{Centrality dependence of pseudorapidity distribution 
with PHOBOS data \cite{PHOBOS_eta}. Impact parameters in the hydrodynamic 
model are 2.4, 4.5, 6.3, 7.5 fm from central to peripheral collisions.     
}
\label{Fig-hydro_eta_cent}
\end{figure}

Figure \ref{Fig-hydro_eta_cent} shows the centrality dependence of the 
pseudorapidity distribution of charged hadrons compared to PHOBOS data   
\cite{PHOBOS_eta}. 
The parameters of our initial condition in the longitudinal direction 
($\sigma_\eta$, $\eta_0$) have been determined by fitting
the data for the most central collisions, no additional
parameter is needed to fix the initial conditions 
for non-central collisions, since they result solely from
the collision geometry. 
The impact parameters for our calculation 
are set to $b=$2.4, 4.5, 6.3 and 7.9 fm   
corresponding to 0-6 \%, 6-15 \%, 15-25 \% and 25-35 \% centrality, 
respectively \cite{PHOBOS_eta}. 
We find good agreement of our results with the experimental 
data not only at mid rapidity 
but also at forward and backward rapidities,     
suggesting that the pure 3-D hydrodynamic model can  
explain the charged hadron multiplicity distribution in a wide range 
of rapidity.          
This observation is consistent with results of other pure hydrodynamic 
models \cite{Hi01,HiTs02,Morita} utilizing different computational
methods.
However, as we show it later, the hydrodynamic calculation overestimates 
the elliptic flow at forward/backward rapidity, indicating that
the rapidity distribution is rather insensitive to the details of the
expansion dynamics.
Having fixed all parameters of our initial condition utilizing 
the $P_T$ spectra and (pseudo)rapidity distributions in the most central
collisions
we now apply our hydrodynamic model to other 
physical observables using these parameters.      

\begin{figure}[tbh]
\begin{minipage}[t]{80mm}
\includegraphics[width=0.9\linewidth]{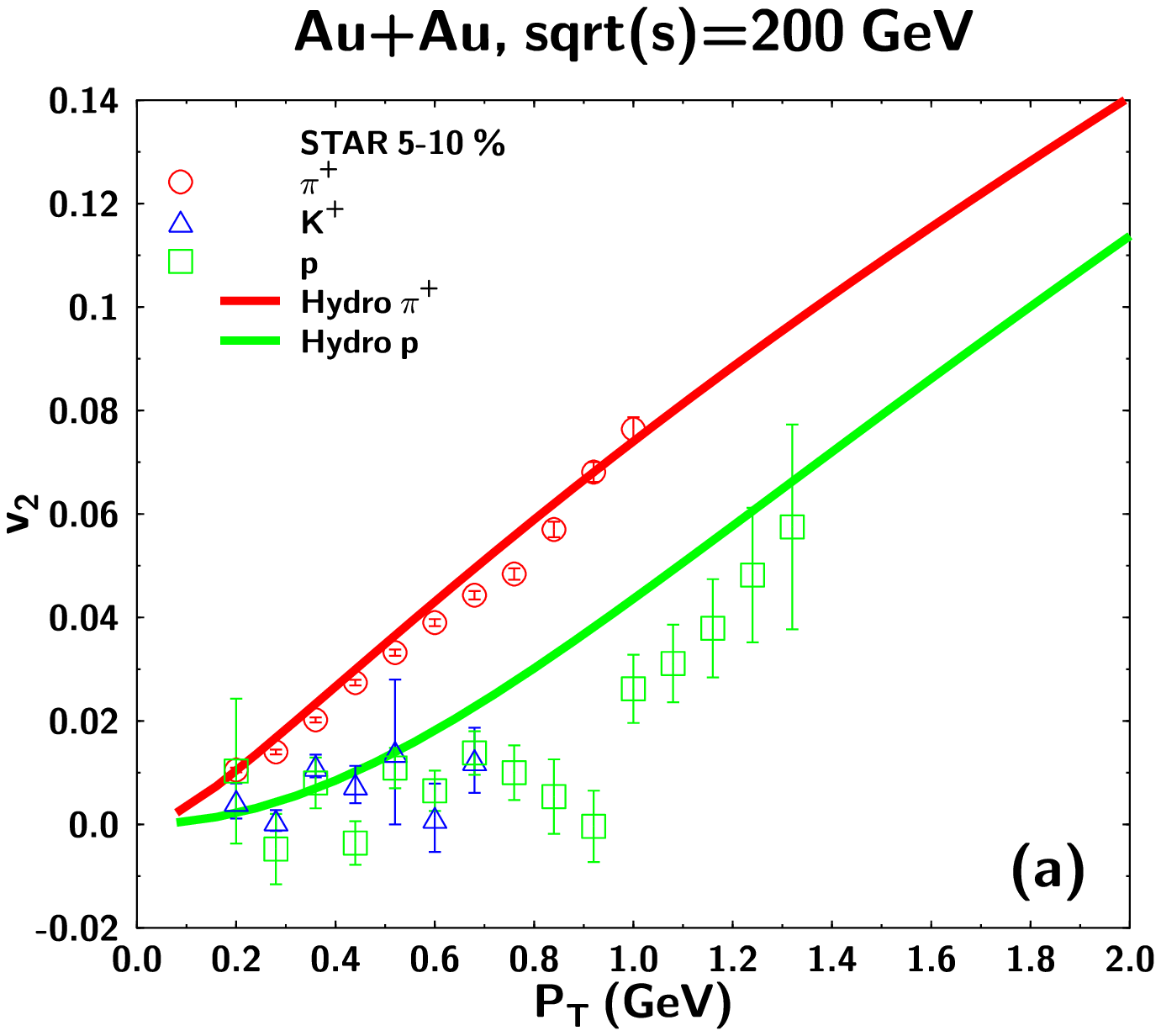}
\end{minipage}
\hspace{3mm}
\begin{minipage}[t]{80mm}
\includegraphics[width=0.9\linewidth]{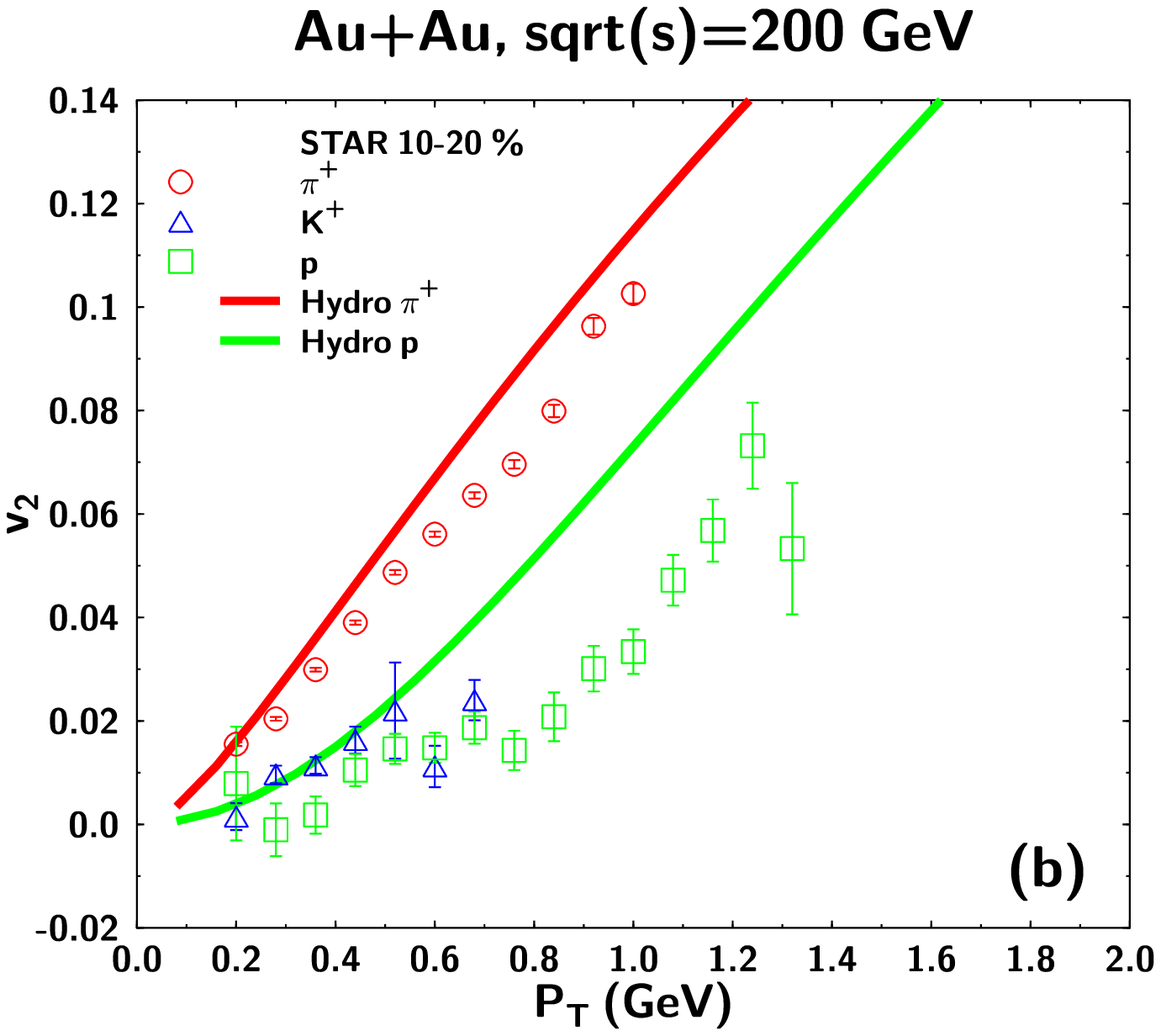}
\end{minipage}
\caption{Elliptic flow as a function of $P_T$ with STAR 
data \cite{STAR_v2} in centrality 5-10 \% (a) and in centrality 10-20 \% (b).    
Impact parameters in hydrodynamic calculations are 4.5 fm (a) and 6.3 fm (b).  
}
\label{Fig-hydro_v2_pt}
\end{figure}

The elliptic flows as a function of $P_T$ in 5-10 \% 
and 10-20 \% most central collisions are shown in 
Fig. \ref{Fig-hydro_v2_pt} (a) and (b) 
together with STAR data \cite{STAR_v2} at mid rapidity. 
We set the impact parameter in our hydrodynamic calculation 
to 4.5 fm (6.3 fm) 
for the 5-10 \% (10-20 \%) data. 
In the case of the 5-10 \% most central events , 
we obtain reasonable agreement to the data
for $\pi$ and $p$. On the other hand, in the 
10-20 \% centrality bin our hydrodynamical calculation 
overpredicts the elliptic flow compared to experimental data. 
Especially for protons, the deviation between calculation and experimental 
data is fairly large -- this trend has already been observed in 
\cite{Hirano:2002ds}.

\begin{figure}[tbh]
\includegraphics[width=0.9\linewidth]{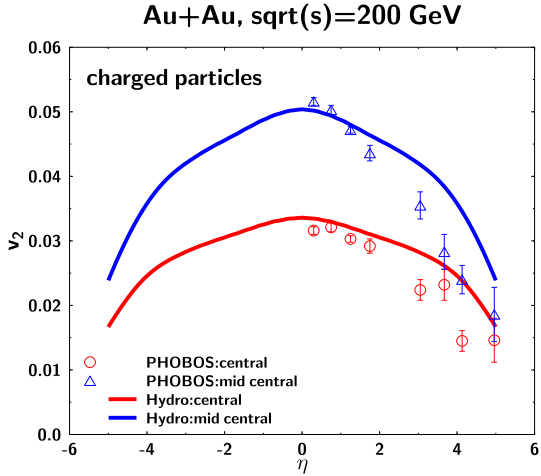}
\caption{Elliptic flow as a function of $\eta$ with PHOBOS 
experimental data \cite{PHOBOS_v2_eta} for central (3-15 \%) and mid 
central collisions (15-25 \%). 
Impact parameters are set to 4.5 (central) and 6.3 (mid central) fm, 
respectively. }  
\label{Fig-hydro_v2_eta}
\end{figure}

Figure \ref{Fig-hydro_v2_eta} shows the elliptic flow as a function of 
$\eta$ in central (3-15 \%) and mid central collisions (15-25 \%). 
In both cases our hydrodynamic   
model calculations overestimate the elliptic flow at forward and backward rapidities, 
similar to results shown in
Ref.~\cite{HiTs02}.    
At large forward and backward rapidities the assumptions  
of a perfect hydrodynamic model such as local equilibrium, 
vanishing mean free path and neglect of viscosity effect apparently 
are no longer valid. 
The deviations at forward and backward rapidities between experimental 
data and calculated results increase with the impact parameter, indicating
a decrease of the volume in which
hydrodynamic limit is achieved.


Summarizing this section, we have applied our ideal 3D RFD model
to Au+Au collisions at the top RHIC energy. A set of parameters
for the initial conditions has been determined which allows for
the simultaneous description of $\pi$, $K$ and $p$ $P_T$ spectra,
the charged hadron rapidity distribution and 
the $P_T$ as well as rapidity dependence of the elliptic flow
coefficient $v_2$ for $\pi$ and $p$. 
Without any additional parameters the 3D RFD model
is capable of describing the centrality dependence of the $P_T$ spectra
and charged hadron rapidity distribution as well.

However, our analysis also has demonstrated a couple of deficits
of the ideal 3D RFD approach -- many of which are already well-known
and have been discussed in the literature before \cite{Hirano:2002ds,
Kolb:2003dz}:
\begin{itemize}
\item the centrality dependence of the elliptic flow coefficient
	$v_2$ as a function of $P_T$ is not well described,
\item the width of the $v_2$ vs. $\eta$ distribution is too broad
	compared to data,
\item  $p$ and multistrange particle spectra need to be normalized 
       by hand in order to account for the separation of chemical 
       and kinetic freeze-out,
\item the hydrodynamic approach is only of limited applicability 
      for small systems (i.e. large impact parameters) and
      large $P_T$ (hard physics).
\end{itemize}

In the following section, we shall 
use the 3D RFD calculation as a baseline to determine the effects
of an improved treatment of the hadronic phase in the framework
of the hydro+micro approach.



\subsection{Application of the hydro+micro approach}
\label{subsec-hybrid}
As in the previous subsection, we first determine 
the parameters of the initial 
conditions in Sec.~\ref{sec-IC} by fitting the 
single particle spectra in the
most central centrality bin.
Table~\ref{table:initial} shows the parameters for both, the pure
3D RFD initial condition as well as the hydro+micro initial condition.
The main difference we find between the two is
in the maximum value of initial energy density.  
The large difference between the two initial energy densities can
be explained by our omission of resonance decays in the pure 
hydrodynamic calculation and to a lesser extent by the dissipative
corrections present in the hydro+micro calculation. Both effects
result in additional particle production, leading to a smaller initial
energy- and entropy-density necessary to describe the final particle
multiplicities.

We set the switching temperature $T_{\rm SW}$ to 
150 MeV. This allows for a brief period of hydrodynamic evolution
in the hadronic phase to account for multi-particle collisions which
can occur at large densities and temperatures in the hadronic
phase close to the phase-boundary \cite{rapp,leupold}. The dependence
of hadronic observables on $T_{\rm SW}$, such as the mean transverse 
momentum $\langle P_T \rangle$ of the different hadron species, has
been investigated in \cite{BaDu00} -- our choice of 150~MeV
conforms to the lower bound of the allowed range for $T_{\rm SW}$.

\begin{figure}[tbh]
\includegraphics[width=0.9\linewidth]{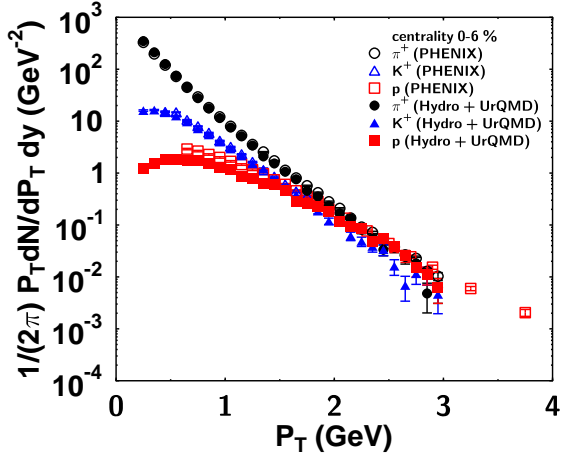}
\caption{$P_T$ spectra for $\pi^+$, $K^+$ and $p$ at central collisions 
with PHENIX data \cite{PHENIX_PT}. The poins are NOT renormalized. 
}
\label{Fig-hu_pt}
\end{figure}

Figure \ref{Fig-hu_pt} shows the $P_T$ spectra of $\pi^+$, $K$ and $p$ at 
$\sqrt{s_{NN}}=200$ GeV central collisions. The most compelling feature,
compared to the pure 3D RFD calculation, is that the hydro+micro 
approach is capable of accounting for the proper normalization of
the spectra for all hadron species
 without any additional  
correction as is performed in the pure hydrodynamic model. 
The introduction of a realistic freezeout process  
provides therefore 
a natural solution to the problem of separating chemical and 
kinetic freeze-out in a pure hydrodynamic approach.

\begin{figure}
\includegraphics[width=0.9\linewidth]{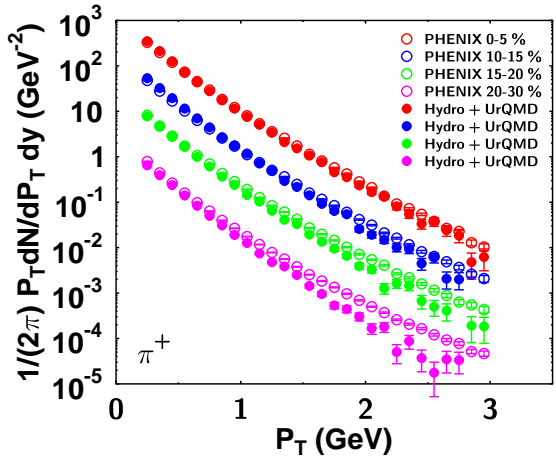}
\caption{Centrality dependence of $P_T$ spectra of $\pi^+$ with 
PHENIX data \cite{PHENIX_PT}. The $P_T$ spectra at 10--15 \%, 15--20 
\% and  20--30 \% are divided by 5, 25 and 200, respectively.
}
\label{Fig-hu_pt_cent}
\end{figure}

In Fig.~\ref{Fig-hu_pt_cent} centrality dependence of $P_T$ spectra of 
$\pi^+$ is shown. The impact parameter for each centrality is determined 
in the same way as in the pure hydrodynamic calculation.
The separation between model results and experiment appears  
at lower transverse momentum in peripheral collisions compared to central 
collisions, just as in the pure hydrodynamic calculation.
The 3D hydro + micro model does not provide any improvement for this behavior, 
since the hard physics high $P_T$  contribution to the spectra
occurs at early reaction times before the system has reached the QGP phase
and is therefore neither included in the pure 3D RFD calculation nor in the
hydro+micro approach.

\begin{figure}[tbh]
\includegraphics[width=0.9\linewidth]{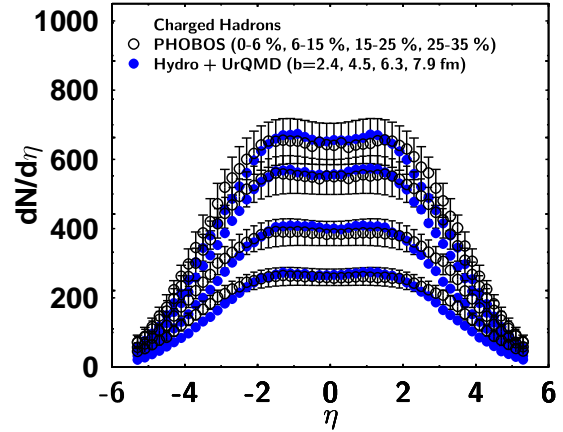}
\caption{Centrality dependence of pseudorapidity distribution of 
charged particles with PHOBOS data \cite{PHOBOS_eta}. 
Impact parameters in a calculation are 2.4 (0-6 \%), 
4.5 (6-10 \%), 6.3 (10-15 \%), 7.5 fm (25-35 \%).   
}
\label{Fig-hu_eta}
\end{figure}

Figure~\ref{Fig-hu_eta} shows the centrality dependence of the  
pseudorapidity distribution of charged hadrons compared to 
PHOBOS data \cite{PHOBOS_eta}.
Solid circles stand for model results and open circles denote
data taken by the PHOBOS collaboration \cite{PHOBOS_eta}.
The impact parameters are set to $b=2.4, 4.5, 6.3, 7.9$ fm for 
0-6 \%, 6-15 \%, 15-25 \% and 25-35 \% centralities, respectively. 
Our results are consistent with experimental data over a wide pseudorapidity 
region. We observe a small deviation around $|\eta|~\sim 3$, which may be 
improved by tuning the parameter $\sigma_\eta$ (here we have chosen the 
same value as for the pure RFD calculation).   
There is no distinct difference between 3-D ideal
RFD model and the hydro + UrQMD model in the centrality dependence of the 
psuedorapidity distribution, indicating that the shape of psuedorapidity 
distribution is insensitive to the detailed microscopic 
reaction dynamics of the hadronic final state.

\begin{figure}[tbh]
\includegraphics[width=0.9\linewidth]{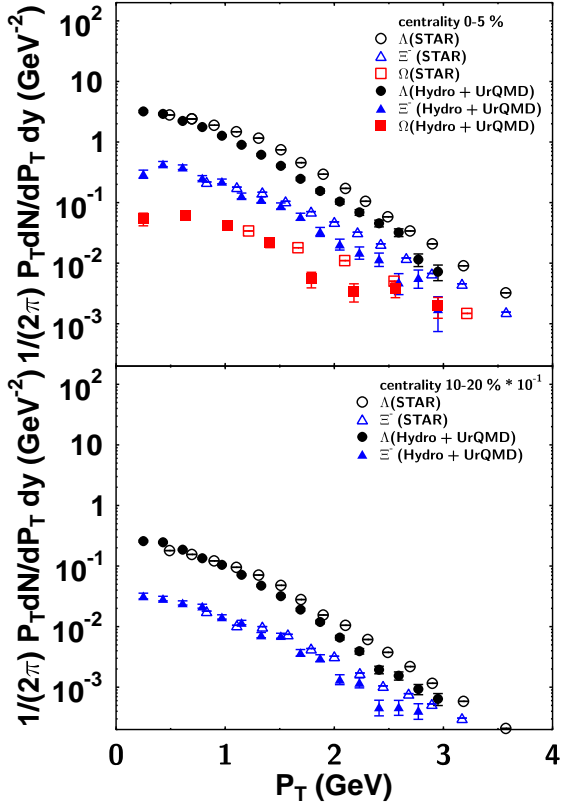}
\caption{$P_T$ spectra of multi-strange particles at centralities 0--5 \%  
and 10--20 \% with STAR data \cite{STAR_strange1}. 
}
\label{Fig-hu_pt_ms}
\end{figure}

In Fig.~\ref{Fig-hu_pt_ms} we analyze the $P_T$ spectra of 
multistrange particles. Our results show good agreement with   
experimental data for $\Lambda$, $\Xi$, $\Omega$ for centralities 0--5 \% 
and 10--20 \%.  In this calculation the additional procedure for normalization 
is not needed.  
Recent experimental results suggest that 
at thermal freezeout multistrange baryons exhibit less transverse flow  
and a higher temperature closer to the chemical freezeout 
temperature compared to non- or single-strange baryons
\cite{STAR_strange1,STAR_strange2}. This behavior can be understood in terms
of the flavor dependence of the hadronic cross section, which decreases
with increasing strangeness content of the hadron. The reduced
cross section of multi-strange baryons leads to a decoupling from the
hadronic medium at an earlier stage of the reaction, allowing them
to provide information on the properties of the hadronizing QGP less
distorted by hadronic final state interactions \cite{vanHecke:1998yu,Dumitru:1999sf}.
It should be noted that the analogous behavior has already been observed
in experiments at the CERN-SPS \cite{sps_multistrange}.
Later in this section we will discuss the reaction dynamics of 
multi-strange baryons  in greater detail by analyzing the baryon 
collision number and freezeout time distributions as well 
as their collision rates.  

\begin{figure}[tbh]
\includegraphics[width=0.9\linewidth]{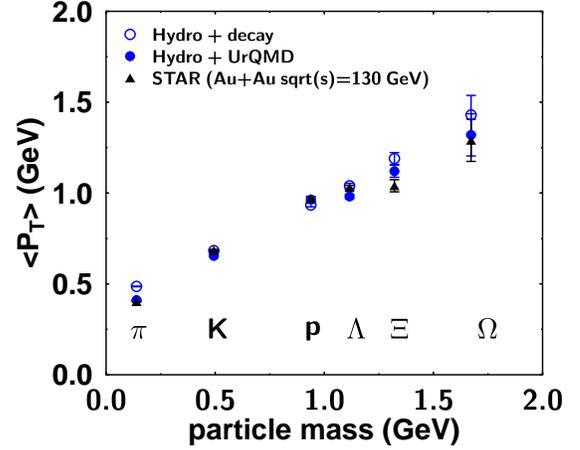}
\caption{Mean $P_T$ as a function of mass. Open circle symbols stand for 
hydro + decay, solid circle symbols stand for hydro + UrQMD and solid 
triangle stand for STAR data (Au+Au $\sqrt{s_{NN}}=130$ GeV).  
}
\label{Fig-mean_PT}
\end{figure}

In Fig.~\ref{Fig-mean_PT} the mean transverse momentum $\langle P_T\rangle $ 
as a function of hadron  
mass  is shown. Open symbols denote the value at $T_{\rm sw}=150$~MeV, corrected
for hadronic decays. Not surprisingly, in 
this case the $\langle P_T\rangle $   follow a straight line, suggesting a 
hydrodynamic expansion.  However if hadronic rescattering 
is taken into account (solid circles) the $\langle P_T\rangle $  
do not follow the straight line any more: 
the $\langle P_T\rangle $ of pions is actually reduced by hadronic
rescattering (they act as a heat-bath in the collective expansion), 
whereas protons actually pick up additional transverse momentum in the
hadronic phase. RHIC data by the STAR collaboration is shown via
the solid triangles -- overall the proper treatment of hadronic
final state interactions significantly improves the agreement of the
model calculation with the data.

\begin{figure}[tbh]
\includegraphics[width=0.9\linewidth]{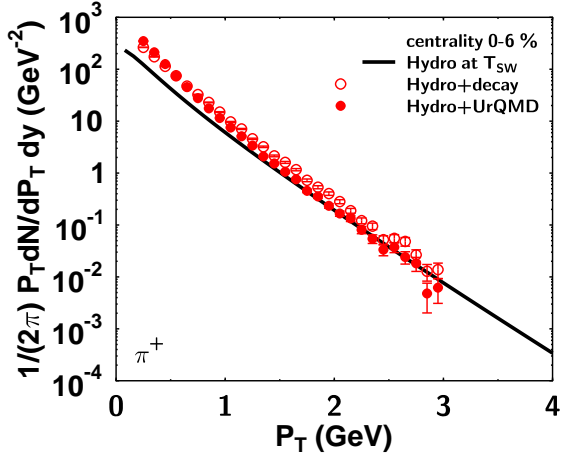}
\caption{$P_T$ spectra of $\pi^+$ from hydro at switching 
temperature (solid line), hydro+decay (open symbols) and 
hydro + UrQMD (solid symbols) at central collision.
}
\label{Fig-hu_final_in}
\end{figure}

Let us now investigate the effect of resonance decays and
hadronic rescattering on the pion and
baryon transverse momentum spectra: Figure~\ref{Fig-hu_final_in}
shows the $P_T$ spectrum for $\pi^+$ at $T_{\rm sw}=150$~MeV
(solid line, uncorrected for resonance decays) 
as well as the final spectrum after hadronic rescattering and
resonance decays, labeled as 
{\em Hydro+UrQMD}
(solid symbols). 
In addition the open symbols denote a calculation
with the resonance decay correction
performed at $T_{\rm sw}$, which we label as {\em Hydro+decay}.
The difference between the solid line and open symbols therefore
directly quantifies the effect of resonance decays on the spectrum,
which is most dominant in the low transverse 
momentum region $P_T < 1$ GeV  
Furthermore, the comparison 
between open symbols and solid symbols quantifies the effect of hadronic
rescattering:
pions with $P_T > 1 $ GeV lose momentum via these final state 
interactions, resulting in a steeper slope.

\begin{figure}[tbh]
\includegraphics[width=0.9\linewidth]{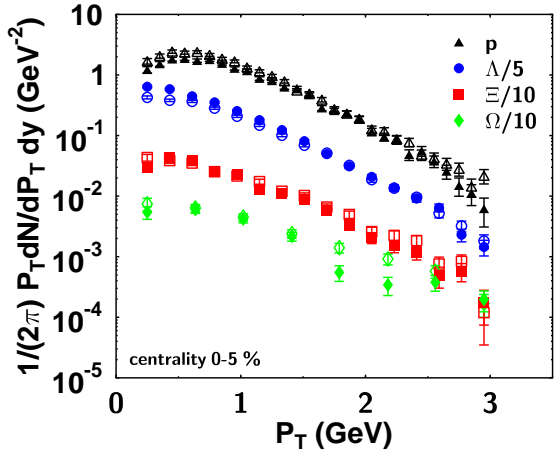}
\caption{$P_T$ spectra of baryons from hydro+decay (open symbols) and 
hydro + UrQMD (solid symbols)  at centrality 0--5 \%.
}
\label{Fig-hu_final_baryon}
\end{figure}

Figure \ref{Fig-hu_final_baryon} shows a likewise analysis for baryons
(p, $\Lambda$, $\Xi$ and $\Omega$).
We note that in contrast to SPS energies \cite{BaDu00}, there is very
little effect on the spectra due to hadronic rescattering - even for
protons which rescatter 8 -- 10 times at mid-rapidity. Only at low
transverse momenta the multiple scatterings of the protons (predominantly
with pions) manifests itself in
a slight flattening of the $P_T$ distribution of
the protons, giving rise to a slight increase in their radial flow.
This phenomenon has been discussed in Ref. \cite{BaDu00} 
and is commonly referred to as `pion wind'. 

\begin{figure}[tbh]
\includegraphics[width=0.9\linewidth]{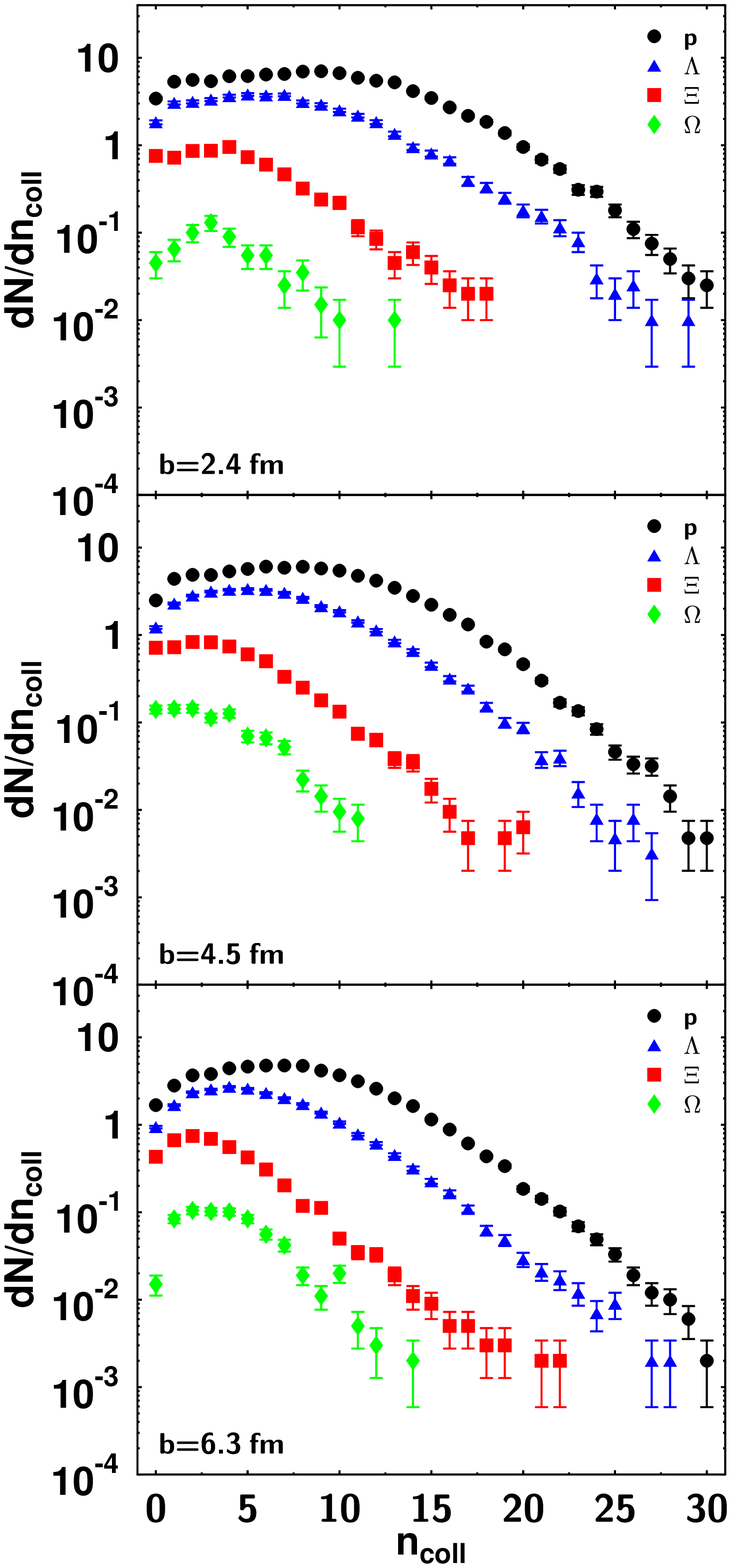}
\caption{The impact parameter dependence of distribution of 
number of collisions for $p$, $\Lambda$, $\Xi$, $\Omega$. 
}
\label{Fig-hu_n_coll}
\end{figure}

\begin{figure}[tbh]
\includegraphics[width=0.9\linewidth]{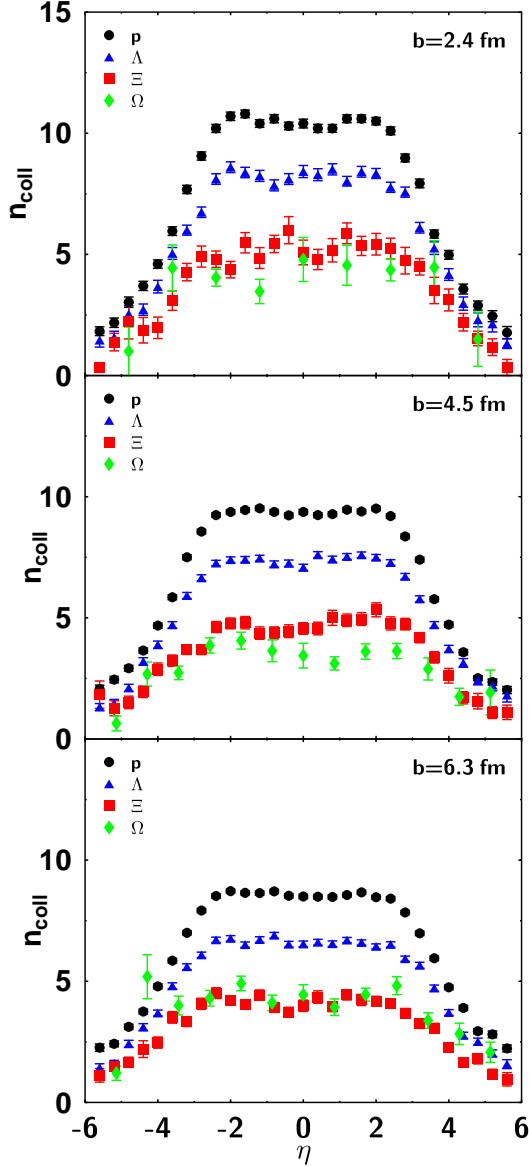}
\caption{The impact parameter dependence of distribution of 
number of collisions as a function of $\eta$ 
for $p$, $\Lambda$, $\Xi$ and $\Omega$.
}
\label{Fig-hu_n_coll_eta}
\end{figure}

Figure \ref{Fig-hu_n_coll} shows the distribution of the number of collisions 
which particles suffer in the hadronic phase at    
$b=2.4$, 4.5, 6.3 fm.  The distributions for protons and lambdas are very broad,
indicating a large amount of rescattering taking place in the hadronic phase.
Around mid-rapidity protons rescatter on average 10 times for central and semi-central
impact parameters and lambdas rescatter 7-8 times (See also Fig.~\ref{Fig-hu_n_coll_eta}.).  
However, the average number of collision of multistrange baryons is less than half of
those for protons and $\Lambda$'s, as is to be expected due to the decrease
of the hadronic cross section with increasing strangeness content of the hadron.
We note that the collision number distributions change significantly
as a function of centrality -- for large impact parameters the high-$n_{\rm coll}$
tail of the distribution exhibits a much steeper drop as a function of $n_{\rm coll}$.

The pseudo-rapidity dependence of the number of hadronic rescatterings
for different baryon species is analyzed in  
Fig.~\ref{Fig-hu_n_coll_eta}, which shows the number of 
collisions of $p$, $\Lambda$, $\Xi$ and $\Omega$ as a function
of $\eta$ at $b=2.4$, 4.5 and 
6.3 fm. The distributions appear to be
similar to that of the particle yield psuedorapidity distribution. 
At midrapidity we find a plateau region extending from $\eta=-3$ to 3, followed
by a steep drop-off to
forward and backward rapidities.
The flavor dependence of the average collision numbers is again clearly seen,
even though we would like to point out that the shapes of the different
distributions is very similar. The large plateau region indicates the 
rapidity domain in which {\em interacting} matter can be found and in which
the application of thermodynamic concepts is viable.

\begin{figure}[tbh]
\includegraphics[width=0.9\linewidth]{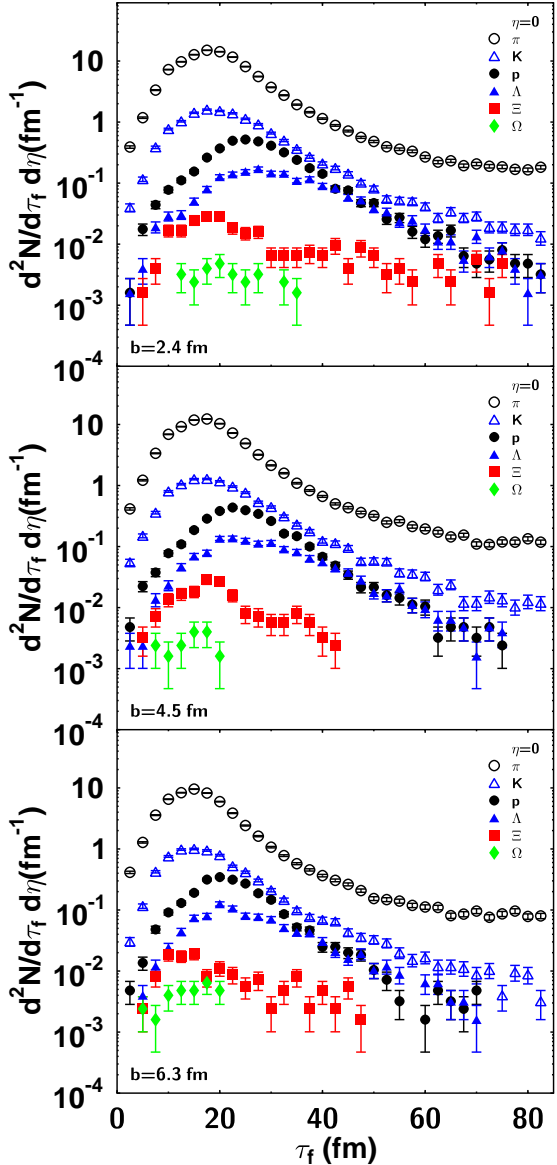}
\caption{The freezeout time distribution of $\pi$, $K$, $p$, $\Lambda$, 
$\Xi$ and $\Omega$ at mid rapidity for $b=2.4$, 4.5, 6.3 fm.  
 }
\label{Fig-hu_tauf}
\end{figure}

Figure \ref{Fig-hu_tauf} depicts the freezeout time distribution 
of $\pi$, $K$, $p$, $\Lambda$, $\Xi$ and $\Omega$ at midrapidity for 
 $b=2.4$, 4.5, 6.3 fm.
The freezeout time distribution of $\pi$ and $K$ has a peak around 
$\tau \sim 19$ ($< 20$ fm), while the peak of $p$  and $\Lambda$ freezeout 
time distribution is shifted to later times. We also note that the 
freezeout time of multistrange baryons is much smaller than that 
of other particles.  
The  freezeout times are in general determined by the amount of 
hadronic rescattering suffered by the different hadron species -- our
analysis indicates that (a)
hadronic freeze-out is strongly species dependent,
and (b) even for a particular species, the freeze-out distribution
      is broad and it is therefore nearly impossible (or at least
	extremely ambiguous) to define a precise freeze-out
	time for a given hadron species.

This figure again supports the finding that multistrange baryons are 
produced with  few interactions right after hadronization and
freeze-out early \cite{vanHecke:1998yu,Dumitru:1999sf}, indications of which 
have been observed by the STAR collaboration \cite{STAR_strange1,STAR_strange2}.   
Given the broad freeze-out time distributions it is very difficult to
quantify the overall duration of the heavy-ion reaction -- a possible criterion
would be the drop-off of the number of freezing out particles per unit time
and rapidity below 1 -- which would put the overall duration of the reaction
to approximately 30 fm/c.

\begin{figure}[tbh]
\includegraphics[width=0.9\linewidth]{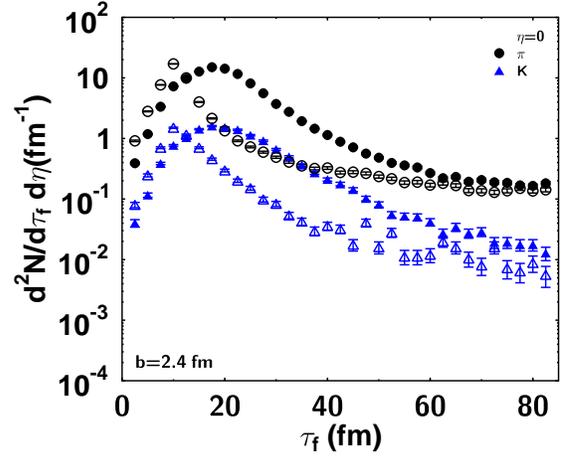}
\caption{The freezeout time distribution of mesons for hydro+decay (open 
symbols) and hydro + UrQMD (solid symbols) at mid rapidity in the 
case of central collision. }
\label{Fig-hu_tauf_m}
\end{figure}

In Fig.~\ref{Fig-hu_tauf_m} we study the effect hadronic rescattering
has on the duration of the freeze-out process by
comparing a calculation terminated
at $T_{\rm sw}$ without hadronic rescattering (open symbols) to one including the full
hadronic final state interactions (solid symbols). 
If we terminate at $T_{\rm sw}$, most hadrons  
freezeout around 10 fm/c, reflecting the lifetime of the deconfined phase
in our calculation (the tails of the distribution stem from the decays
of long-lived resonances).
The inclusion of hadronic rescattering shifts
the peak of the freezeout distribution to larger freezeout 
times ($\tau_f \sim 20-30$), providing us with an estimate on the
lifetime of the hadronic phase around 10-20 fm/c. Note that this estimate
is subject to the same systematic uncertainties discussed previously in 
the context of the overall lifetime of the system. 

\begin{figure}[tbh]
\includegraphics[width=0.9\linewidth]{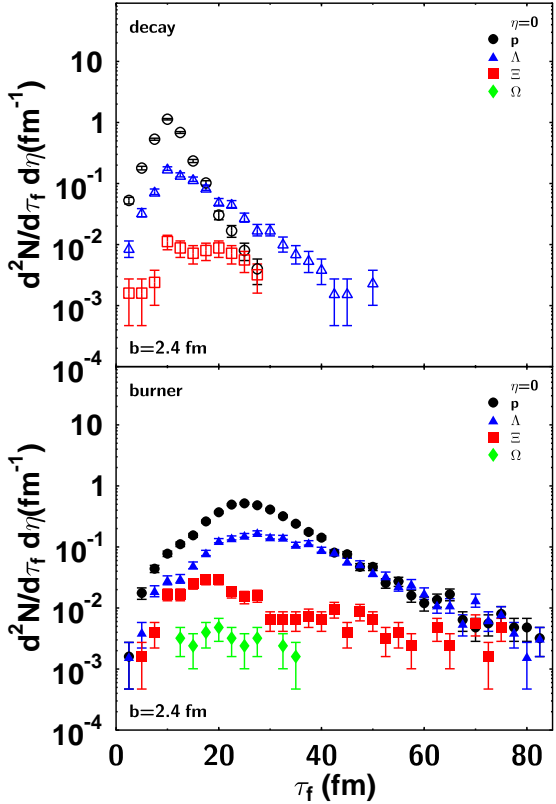}
\caption{The freezeout time distribution of baryons for 
hydro+decay (open symbols, above) and hydro + UrQMD (solid symbols, below) 
at mid rapidity. }
\label{Fig-hu_tauf_b}
\end{figure}

The findings discussed in the context of the previous figure for
pions and kaons are confirmed by analyzing baryons in the same fashion,
which is shown in Fig.~\ref{Fig-hu_tauf_b}: here the top frame
contains the analysis terminated at $T_{\rm sw}$ and the bottom frame
contains the calculation including full hadronic rescattering. 

\begin{figure}[tbh]
\includegraphics[width=0.9\linewidth]{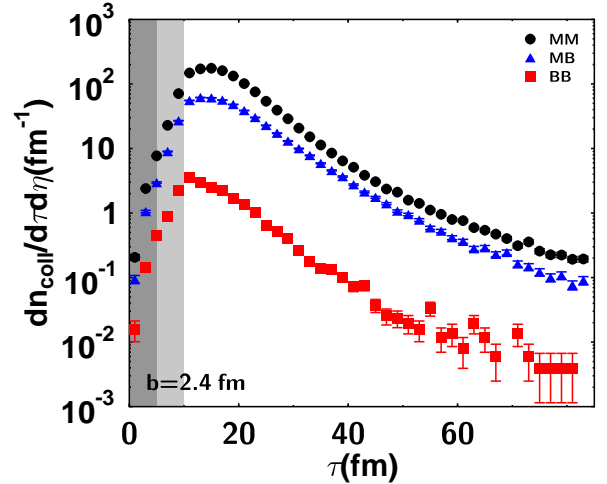}
\caption{Collision rate for meson-meson (MM), meson-baryon (MB) and 
baryon-baryon (BB) at central collisions. The (dark) grey zone stands for 
qgp (mixed) phase, which is determined by the center cell.     
}
 \label{Fig-hu_cr}
 \end{figure}

Figure \ref{Fig-hu_cr} shows collision rates for meson-meson, 
meson-baryon and baryon-baryon collisions at mid-rapidity and 
an impact parameter of $b=2.4$ fm. 
We find that the reaction dynamics in the hadronic phase is dominated
by meson-meson and to a lesser degree by meson-baryon interactions. 
Baryons essentially propagate in a medium dominated by mesons -- a situation
very different from the regime at AGS and lower SPS energies.

\begin{figure}[tbh]
\includegraphics[width=0.9\linewidth]{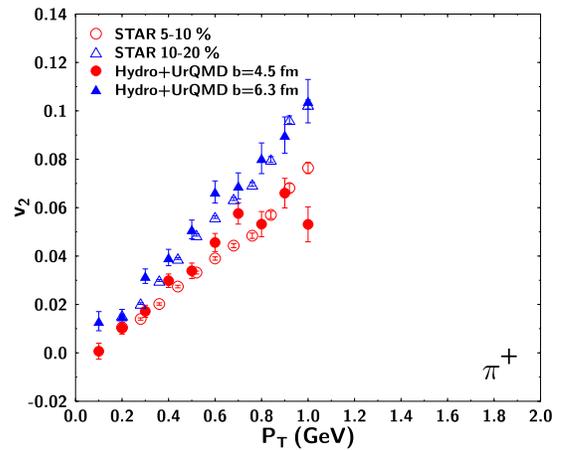}
\caption{Elliptic flow as a function of $P_T$  for $\pi^+$ at 
centralities 5-10 \% and 10-20 \%. Open symbols stand for 
STAR data and solid symbols stands for our results.    
}
\label{Fig-v2_pt_pip}
\end{figure}

Finally, let us move to the analysis of elliptic flow in the hydro+micro
approach: 
First we show the  elliptic flow coefficient $v_2$ as a function of $P_T$ 
for $\pi^+$ at centralities 5-10 \% and 10-20 \%  in Fig.~\ref{Fig-v2_pt_pip}. 
Open symbols stand for experimental data and solid symbols 
represent our calculations. We find that the hydro+micro approach is able
to provide an improved agreement to the data compared to the purely
hydrodynamic calculation (see Fig.~\ref{Fig-v2_pt_pip}  for comparison).
 
\begin{figure}[tbh]
\includegraphics[width=0.9\linewidth]{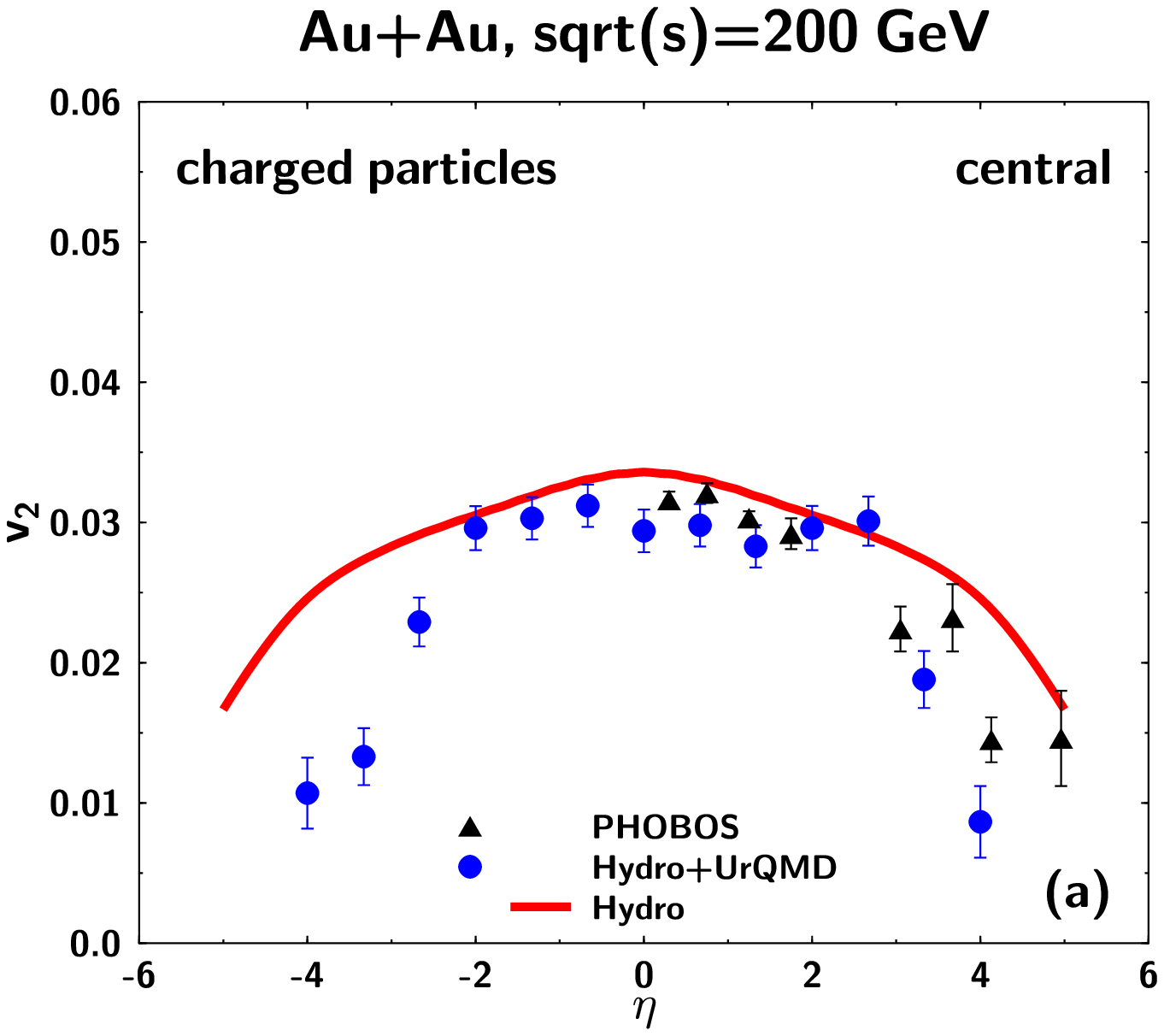}
\includegraphics[width=0.9\linewidth]{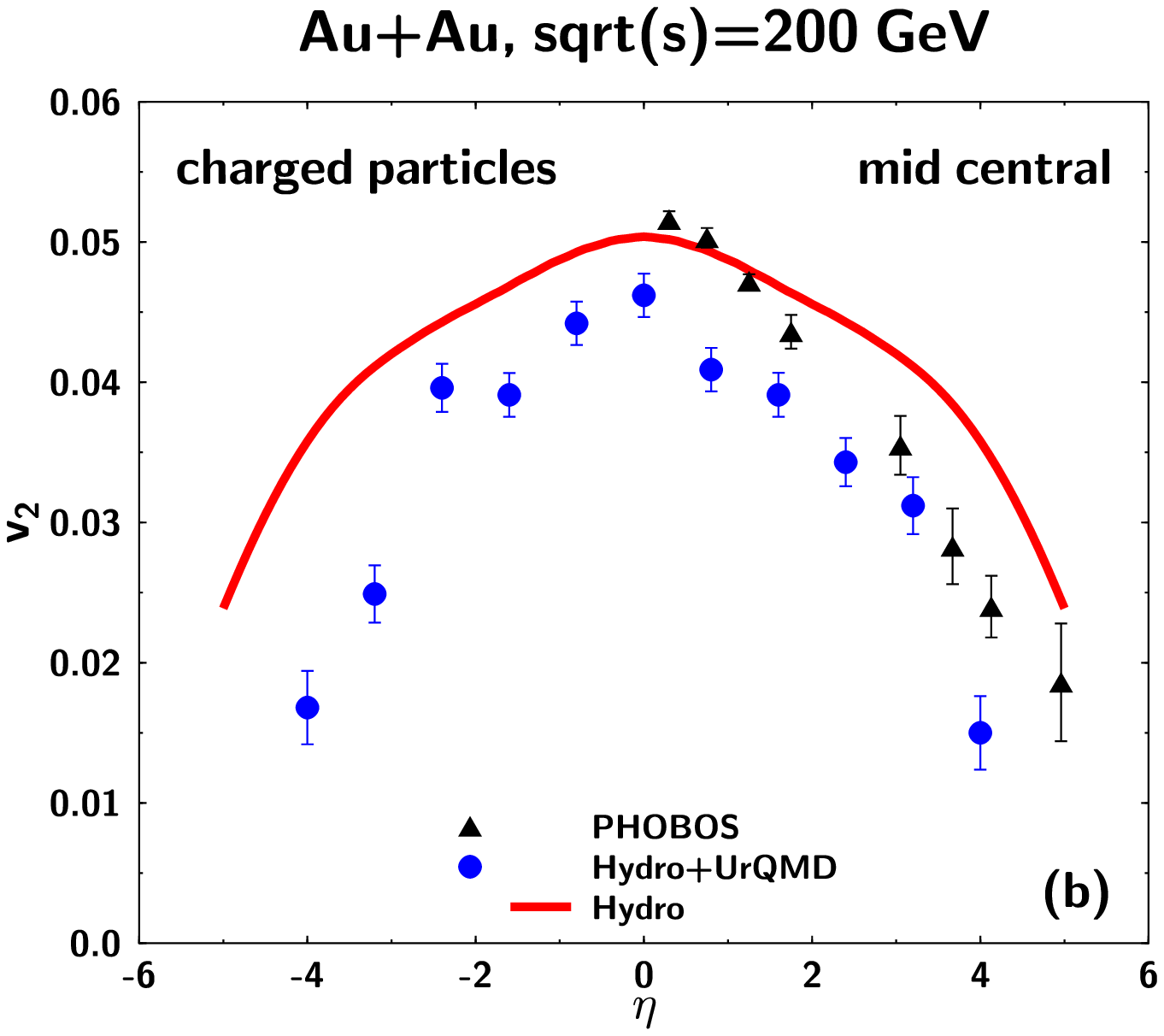}
\caption{Elliptic flow as a function of $\eta$ of charged particles at  
central collisions (a) and mid central collisions (b).    
Impact parameters in the calculation are set to 4.5 fm (a) and 6.3 fm (b), 
respectively. The solid line stands for the 3-D pure hydrodynamic 
calculation and solid circles stand for results by hydro+UrQMD with 
PHOBOS data~\cite{PHOBOS_v2_eta}. 
}
\label{Fig-hu_v2_eta}
\end{figure}

Figure~\ref{Fig-hu_v2_eta} shows the elliptic flow coefficient $v_2$ as a function of 
pseudo-rapidity $\eta$ for charged particles 
in central (3-15 \%) (a) and mid central 
(15-25 \%) (b) collisions.      
The solid line stands for a purely hydrodynamic calculation 
whereas the solid circles denote
the hydro+micro approach and the solid triangles 
represent PHOBOS data~\cite{PHOBOS_v2_eta}.  
We find that the dissipative effects contained in the hydro+micro
approach significantly alter the shape of the $v_2$ vs.~$\eta$ curve,
providing a far better agreement to the data -- in particular for 
pseudo-rapidities away from mid-rapidity.
The analysis shows that apparently dissipative effects increase towards
projectile and target rapidities, where the assumptions of ideal fluid
dynamics break down earlier. Our analysis confirms the findings 
of~\cite{Hirano:2005xf}, who in addition also studied the effect
of a Color-Glass initial condition on the elliptic flow rapidity dependence.

\begin{figure}[tbh]
\includegraphics[width=0.9\linewidth]{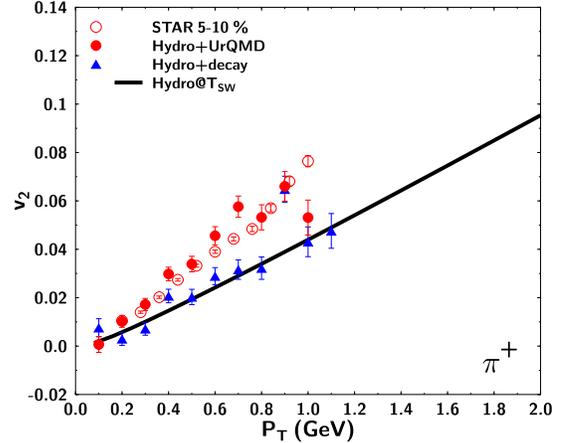}
\caption{Elliptic flow as a function of $P_T$ of $\pi^+$ at 
centrality 5-10 \% from pure hydro at the switching temperature 
(solid line), hydro + decay (solid triangles) and hydro + UrQMD (solid 
circles). Open symbols stand for experimental data from STAR. 
}
\label{Fig-hu_v2_pt_pip}
\end{figure}

\begin{figure}
\includegraphics[width=0.9\linewidth]{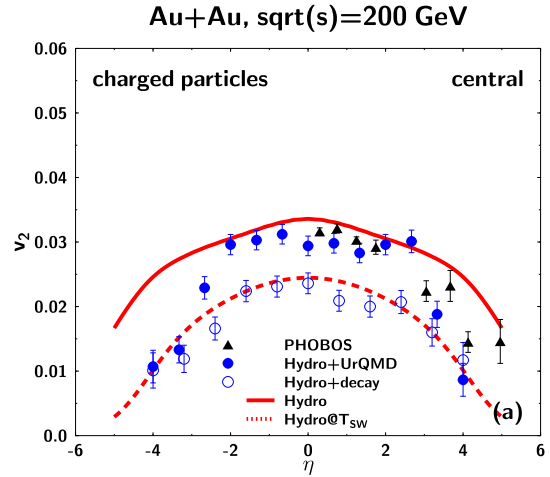}
\caption{Elliptic flow as a function of $\eta$ of charged particles. 
Solid line and dashed lines stand for pure hydro calculation at the freezeout 
temperature (110 MeV) and the switching temperature, respectively. 
Experimental data by PHOBOS is shown with solid triangles. 
Open (Solid) circles stand for hydro + decay (hydro + UrQMD).  
}
\label{Fig-hu_v2_eta_decay}
\end{figure}

The question of how much  elliptic flow develops during
the deconfined phase vs. the hadronic phase is investigated in
Figs.~\ref{Fig-hu_v2_pt_pip} and~\ref{Fig-hu_v2_eta_decay}. 
It has been pointed out repeatedly, both in the framework
of microscopic \cite{Sorge:1996pc}
as well as as hydrodynamic analysis \cite{KoSoHe00} that 
elliptic flow develops early on during the deconfined phase
of the reaction and thus serves as a sensitive tool to the equation
of state of QCD matter. This picture has been confirmed by the experimentally
observed
quark number scaling $v^{h}_2 \sim  nv_2 (1/n P_T)$ 
of elliptic flow at intermediate transverse momenta 
$P_T$ \cite{Adams:2003am,Molnar:2003ff}.

In Fig.~\ref{Fig-hu_v2_pt_pip} we plot elliptic flow $v_2$ as a function of $P_T$. 
The solid line stands for the pure hydro calculation, terminated at the switching 
temperature $T_{\rm sw}$ and solid circles denote the full
hydro+micro calculation. We find that the QGP contribution to the 
elliptic flow depends on the transverse momentum -- for low $P_T$ nearly
100\% of the elliptic flow is created in the QGP phase of the reaction,
whereas the hadronic phase contribution increases to 25\% at a $P_T$ 
of 1~GeV/c. 
%

Figure \ref{Fig-hu_v2_eta_decay} shows the elliptic flow as a 
function of $\eta$: the pure hydrodynamic calculation is shown by 
the solid curve, the hydrodynamic contribution at $T_{\rm sw}$ is denoted
by the dashed line and the full hydro+micro calculation is given by
the solid circles, together with PHOBOS data (solid triangles). 
The shape of  the elliptic flow in the pure hydrodynamic calculation at 
$T_{\rm sw}$ is quite different from that of the full
hydrodynamic one terminated at a freeze-out temperature of 110~MeV. 
Apparently the slight bump at forward and backward rapidities observed
in the full hydrodynamic calculation develops first in the later hadronic
phase, since it is not observed in the calculation terminated at 
$T_{\rm sw}$.
Evolving the hadronic phase in the hydro+micro approach will increase
the elliptic flow at central rapidities, but not in the projectile and 
target rapidity domains.
As a result, the elliptic flow calculation in the hydro+micro approach
is closer to the experimental data when compared to the pure hydrodynamic calculation.

\section{Summary and outlook}

In summary, 
we have introduced a hybrid macroscopic/microscopic transport approach,
combining a newly developed relativistic 3+1 dimensional
hydrodynamic model for the early deconfined stage of the reaction 
and the hadronization process
with a microscopic non-equilibrium model for the later hadronic
stage at which the hydrodynamic equilibrium assumptions are not valid anymore.
Within this approach we have dynamically 
calculated the
freezeout of the hadronic system,
accounting for the collective flow on the
hadronization hypersurface generated by the QGP expansion.
We have compared the results of our hybrid model and of a calculation 
utilizing our hydrodynamic model for the full evolution of the reaction to
experimental data. This comparison has allowed us to quantify the strength
of dissipative effects prevalent in the later hadronic phase of the reaction,
which cannot be properly treated in the framework of ideal hydrodynamics.

Overall, the improved treatment of the hadronic phase provides a far better
agreement between transport calculation and data, in particular concerning
the flavor dependence of radial flow observables and the collective behavior 
of matter at forward/backward rapidities. We find that the hadronic
phase of the heavy-ion reaction at top RHIC energy is of significant duration
(at least 10~fm/c) and that hadronic freeze-out is a continuous process,
strongly depending on hadron flavor and momenta.

With this work we have established a base-line -- both for the regular
3+1 dimensional hydrodynamic model as well as for the hybrid hydro+micro
approach. In forthcoming publications we shall expand on this baseline by
investigating the effects of a realistic lattice-QCD motivated equation of state 
containing a tri-critical point and by performing an analysis of two particle
correlations (HBT interferometry). We also plan to use our model as the medium
for the propagation of jets and heavy quarks and to study the modification of 
our medium due to the passage of these hard probes.

\acknowledgements
We would like to thank Berndt Mueller and Ulrich Heinz 
for many valuable discussions and Berndt Mueller for the  
careful reading of this manuscript. 
This work was supported by an Outstanding Junior Investigator Award
under grant number DE-FG02-03ER41239 and a 
JSPS (Japan Society for the Promotion of Science) Postdoctoral 
Fellowship for Research Abroad. 

\section{appendix}
Equation (\ref{Eq-rhydro}) in $(\tau,x,y,\eta)$ is written in the 
following explicit way,      
\begin{widetext}
\begin{eqnarray}
 &
\left ( 
\begin{array}{ccccc}
\tg^2 \tvx & \tg^2 \tvy & \tg^2 \tvh & \iw \dET & \iw  \dEm \\
\tg^2      &  0         &  0         & \iw \tvx \dPT & \iw \tvx \dPm \\
0          & \tg^2      &  0         & \iw \tvy \dPT & \iw \tvy \dPm \\
0          & 0          &  \tg^2     & \iw \tvh \dPT & \iw \tvh \dPm \\
n_B \tg^2 \tvx  & n_B \tg^2 \tvy     &  n_B \tg^2 \tvy      & \dnT & \dnm \\
\end{array}
\right ) 
\partial_\tau 
\left (
\begin{array}{c}
\tvx \\ \tvy \\ \tvh \\ T \\ \mu 
\end{array}
\right )
  &  \nonumber \\
+ & 
\left (
\begin{array}{ccccc}
\tg^2 \tvx^2 + 1 & \tg^2 \tvx \tvy & \tg^2 \tvx \tvh & 
\iw \tvx \dET  & \iw \tvx \dEm \\
\tg^2 \tvx & 0 & 0 & \iw \dPT & \iw \dPm \\
0 & \tg^2 \tvx  & 0 & 0 & 0 \\
0 & 0 & \tg^2 \tvx & 0 & 0 \\
n_B (\tg^2 \tvx^2 + 1) & n_B \tg^2 \tvx \tvy & n_B \tg^2 \tvx \tvh & 
\tvx \dnT & \tvx \dnm \\
\end{array}
\right )
\partial _x
\left (
\begin{array}{c}
\tvx \\ \tvy \\ \tvh \\T \\ \mu
\end{array}
\right) 
&  \nonumber \\
+ & 
\left ( 
\begin{array}{ccccc}
\tg^2 \tvy \tvx & \tg^2 \tvy^2 + 1 & \tg^2 \tvy \tvh & \iw \tvy \dET & 
\iw \tvy \dEm \\
\tg^2 \tvy & 0 & 0 & 0 & 0 \\
0 & \tg^2 \tvy & 0 & \iw \dPT & \iw \dPm \\
0 & 0 & \tg^2 \tvy & 0 & 0 \\
n_B \tg^2 \tvx \tvy & n_B (\tg^2 \tvy^2 + 1) & n_B \tg^2 \tvy \tvh & 
\tvy \dnT & \tvy \dnm 
\end{array}
\right )
\partial_y
\left (
\begin{array}{c}
\tvx \\ \tvy \\ \tvh \\ T \\ \mu
\end{array}
\right )
& \nonumber \\ 
+  &
\left ( 
\begin{array}{ccccc}
\tg^2 \tvh \tvx & \tg^2 \tvh \tvy & \tg^2 \tvh^2 + 1 & 
\iw \tvh \dET & \iw \tvh \dEm \\
\tg^2 \tvh & 0 & 0 & 0 & 0 \\
0 & \tg^2 \tvh & 0 & 0 & 0 \\
0 & 0 & \tg^2 \tvh & \iw \dPT & \iw \dPm \\ 
n_B \tg^2 \tvx \tvh & n_B \tg^2 \tvh \tvy & n_B (\tg^2 \tvh^2 + 1) & 
\tvh \dnT & \tvh \dnm 
\end{array}
\right )
\frac{1}{\tau}\partial_\eta
\left (
\begin{array}{c}
\tvx \\ \tvy \\ \tvh \\ T\\ \mu
\end{array}
\right )
&
\nonumber \\
 + & 
\left( 
\begin{array}{c}
  \frac{1}{\tau} \\
- \frac{1}{\tau}\tvx \tvh^2 \tg^2   \\
- \frac{1}{\tau}\tvy \tvh^2\tg^2     \\ 
- \frac{1}{\tau}(\tg^2 \tvh-\tvh^3 \tg^2)     \\
  \frac{1}{\tau}n_B
\end{array}
\right)
& = 0, 
\label{Eq-rhydro2}
\end{eqnarray}
\end{widetext}
where $\tg=1/\sqrt{1-\tvx^2-\tvy^2-\tvh^2}$, $Y_L=1/2 \ln ((1+v_z)/(1-v_z))$. 
The difference between the relativistic hydrodynamic 
equations in Cartesian coordinates and those in the  
($\tau, x, y, \eta$) coordinates 
is the addition of the fifth term in Eq.~(\ref{Eq-rhydro2}).   


\end{document}